\documentclass[journal]{IEEEtran}
\usepackage{graphicx}
\usepackage{subfigure}
\usepackage{caption2}
\usepackage{fancyhdr}
\usepackage{amssymb}
\usepackage{amsthm}
\usepackage{amsmath}
\usepackage{mathtools}
\usepackage{citesort}
\usepackage[noadjust]{cite}
\usepackage{dsfont}

\begin{document}
\title{Rate Analysis of Two-Receiver MISO Broadcast Channel with Finite Rate Feedback: A Rate-Splitting Approach}

\author{Chenxi Hao, Yueping Wu and Bruno Clerckx % <-this % stops a space
\thanks{The authors are with the Communication and Signal Processing group of Department of Electrical and Electronic Engineering, Imperial College London, email: {chenxi.hao10;yueping.wu;b.clerckx@imperial.ac.uk}. Dr. Bruno Clerckx is also with the School of Electrical Engineering, Korea University. This work was partially supported by the Seventh Framework Programme for Research of the European Commission under grant number HARP-318489.}% <-this % stops a space
}

\maketitle

\begin{abstract}
\boldmath
To enhance the multiplexing gain of two-receiver Multiple-Input-Single-Output Broadcast Channel with imperfect channel state information at the transmitter (CSIT), a class of Rate-Splitting (RS) approaches has been proposed recently, which divides one receiver's message into a common and a private part, and superposes the common message on top of Zero-Forcing precoded private messages. In this paper, with quantized CSIT, we study the ergodic sum rate of two schemes, namely RS-S and RS-ST, where the common message(s) are transmitted via a space and space-time design, respectively. Firstly, we upper-bound the sum rate loss incurred by each scheme relative to Zero-Forcing Beamforming (ZFBF) with perfect CSIT. Secondly, we show that, to maintain a constant sum rate loss, RS-S scheme enables a feedback overhead reduction over ZFBF with quantized CSIT. Such reduction scales logarithmically with the constant rate loss at high Signal-to-Noise-Ratio (SNR). We also find that, compared to RS-S scheme, RS-ST scheme offers a further feedback overhead reduction that scales with the discrepancy between the feedback overhead employed by the two receivers when there are alternating receiver-specific feedback qualities. Finally, simulation results show that both schemes offer a significant SNR gain over conventional single-user/multiuser mode switching when the feedback overhead is fixed.

\end{abstract}
\newtheorem{myprop}{Proposition}
\newtheorem{mytheorem}{Theorem}
\newtheorem{mycoro}{Corollary}
\newtheorem{myasump}{Assumption}
\newtheorem{myremark}{Remark}
\newtheorem{mylemma}{Lemma}
\newtheorem{myobservation}{Observation}
\section{Introduction}\label{sec:Intro}
\begin{table*}[t]
\centering
\renewcommand{\captionfont}{\small}
\subtable[RS-S vs. conventional schemes in the scenario with equal feedback qualities]{
\begin{tabular}{|c|l|}
\hline
SNR gain offered by additional & RS-S: $\frac{3b}{M{-}1}$ dB \\
$b$-bit feedback overhead & ZFBF-RVQ: sum rate saturates \\
& TDMA: negligible gain\\
\hline
Feedback overhead reduction & RS-S vs. ZFBF-RVQ: $(M{-}1){\log}_2\frac{\frac{\delta}{2e}{+}\frac{e}{2}{-}1}{\sqrt{\delta}{-}1}$ bits\\
\hline
\end{tabular}}\\
\subtable[Benefits offered by RS-ST over RS-S in the scenario with alternating receiver-specific feedback qualities]{
\begin{tabular}{|c|c|}
\hline
SNR gain incurred by $\tau$ bits feedback & $3(\frac{\tau}{2(M{-}1)}{-}2)$ dB \\
overhead differences in each channel use\\
\hline
Feedback overhead reduction & $\frac{\tau}{2}{-}2(M{-}1)$ bits\\
\hline
\end{tabular}}
\caption{Highlights of main contributions.}\label{tab:findings}
\end{table*}
In downlink Broadcast Channel (BC), the utilization of multiple antennas at the transmitter offers a higher multiplexing gain, i.e., Degrees-of-Freedom (DoF), and throughput enhancement compared to the single antenna case. However, to realize such benefits, interference mitigation methods are required at the transmitter and their performance strongly relies on highly accurate channel state information at the transmitter (CSIT), which is difficult to attain in practice.

Under a general assumption that the CSIT error decays with the Signal-to-Noise-Ratio (SNR) as ${\rm SNR}^{-\alpha}$, where $\alpha{\in}[0{,}1]$ is termed as the CSIT quality, conventional multiuser transmission strategy, such as ZFBF, achieves the sum DoF $2\alpha$ in the two-receiver MISO BC. Such a sum DoF performance is worse than single-user transmission when $\alpha{\leq}0.5$ and becomes interference limited when $\alpha{=}0$. To enhance the sum DoF performance, a Rate-Splitting (RS) approach was firstly introduced in literature\footnote{Literature \cite{Ges12} finds the optimal DoF region of two-receiver MISO BC with a mixture of imperfect current CSIT and perfect delayed CSIT. However, one of the corner point of the DoF region can be achieved with the Rate-Splitting approach, which does not rely on perfect delayed CSIT and is applicable to the scenario with only imperfect current CSIT.} \cite[Lemma 2]{Ges12}. In this scheme, the message intended for one receiver is split into a private part and a common part. The private message and the other receiver's (private) message are transmitted via ZFBF using a fraction of the total power, while the common message is superposed on top of the ZF-precoded private messages using the remaining power. Each receiver firstly decodes the common message, and secondly decodes the desired private message via Successive Interference Cancelation (SIC). Since the achievability of the DoF of the common message does not rely on interference nulling, RS-S is more robust to the CSIT error, especially when the CSIT error decreases slowly with SNR (i.e., small value of $\alpha$). When the two receivers have equal CSIT qualities, i.e., $\alpha$, the resultant sum DoF is $1{+}\alpha$, which is larger than $2\alpha$ that is achieved with ZFBF. Based on an assumption of real input and channel vector, the optimality of this sum DoF performance is shown by the recent work\footnote{Literature \cite{Davoodi14} focuses on the scenario where the CSIT of one receiver is perfect, whose sum DoF can be considered as an upper-bound of the sum DoF in the scenario considered in Lemma 2 \cite{Ges12}. As the achievable sum DoF $1{+}\alpha$ is consistent with the upper-bound found in \cite{Davoodi14}, we can say the sum DoF $1{+}\alpha$ is optimal.} \cite{Davoodi14}. This scheme is termed as RS-S scheme in this paper, as the common message is transmitted via a space design. %Similar discussions can be found in \cite{pimrc2013,Elia13,IT_MISO_BC,Jinyuan_evolving_misobc}.

Moreover, in a scenario with alternating receiver-specific CSIT qualities, namely the CSIT quality of receiver 1 (Rx1) and receiver 2 (Rx2) in channel use 1 is $\beta$ and $\alpha$ respectively while the CSIT quality of Rx1 and Rx2 in channel use 2 is $\alpha$ and $\beta$ respectively, performing RS-S scheme in each individual slot/subband leads to a sum DoF of $1{+}\min\{\alpha{,}\beta\}$. This result is unsatisfactory due to its inefficient use of the alternating CSIT qualities. To enhance the DoF performance in this scenario, \cite{Tandon12,Elia13} proposed a more advanced scheme. Compared to the RS-S scheme, this scheme transmits an additional common message across the two channel uses and is denoted here as the RS with space-time design (RS-ST), which results in a sum DoF $1{+}\frac{\alpha{+}\beta}{2}$. The discussion in various CSIT uncertainty scenarios are reported in \cite{pimrc2013,Elia13,IT_MISO_BC,Jinyuan_evolving_misobc}.

In contrast to \cite{Ekrem10,Ekrem12,Weigarten06common} which studied the capacity region of BC with common messages that carry information intended for both receivers, the common messages considered in the RS approaches consist of the \emph{common parts} of the receivers' messages. Although they should be decoded by both receivers, they carry messages to a single-receiver. Nonetheless, all the aforementioned works \cite{Ges12,Davoodi14,Tandon12,Elia13,pimrc2013,Elia13,IT_MISO_BC,Jinyuan_evolving_misobc} focus on a DoF analysis, leaving aside the question of how the Rate-Splitting approach can benefit the ergodic sum rate performance. Tackling such a question is more interesting and meaningful as it sheds light on the usefulness of the information-theoretic works in a practical multiuser MISO system.

In the context of ergodic sum rate analysis in a multiuser MISO BC with imperfect CSIT, there have been extensive works under the finite rate feedback model, where each receiver has to quantize its CSI using a finite number of bits and report it to the transmitter. The impact of the quantized CSIT on the throughput performance of a single-user system was reported in \cite{Love05grassSM,Love07BasicRVQ,Mukka03BF,love04}, while \cite{Jin06,Carie10} focused on a multiuser MISO BC and evaluated the per-receiver rate performance achieved via conventional ZFBF with quantized CSIT. The key finding of \cite{Jin06} reveals that to achieve a constant rate gap relative to ZFBF with perfect CSIT, the number of feedback bits needs to scale with the SNR and the number of transmit antennas. Focusing on Tomlinson-Harashima precoding (THP) with quantized-CSIT, a similar scaling law of the number of feedback bits to achieve a certain maximum allowable rate loss relative to THP with perfect CSIT was found in \cite{Sun13THP}. Note that all these works considered conventional multiuser transmission strategies without integrating common messages.

To the best of our knowledge, the sum rate performance achieved with the aforementioned RS-S and RS-ST scheme in the presence of quantized CSIT remains to be investigated. Hence, in this paper, our objective is to find the benefits of 1) splitting the messages into a common and a private part, and 2) performing RS-ST rather than RS-S when there are alternating receiver-specific feedback qualities, in terms of sum rate performance and feedback overhead reduction compared to the findings in \cite{Jin06}. More specifically, we consider 1) a two-receiver MISO BC, where the number of transmit antennas is greater than or equal to $2$, 2) Random Vector Quantization (RVQ) codebook is employed to quantize the channel vectors, and 3) linear precoders are used in both schemes. Note that in the companion papers \cite{HamdiRsconf1,HamdiRsconf2,HamdiJMB}, RS approach is investigated from a robust beamforming design perspective, which differs from this paper that focuses on a rate analysis. Table \ref{tab:findings} briefly summarizes the main findings, where $M$ refers to the number of transmit antennas and ${\log}_2\delta$ bps/Hz represents a maximum allowable rate loss relative to ZFBF with perfect CSIT. To be more specific, we highlight the main contributions as follows.
\begin{itemize}
\item We derive an upper-bound on the sum rate loss incurred by the RS-S scheme relative to ZFBF with perfect CSIT in the scenario where the two receivers have equal feedback qualities. When the number of feedback bits does not change with SNR, the upper-bound indicates that a $b$-bit increase of the feedback overhead leads to a $\frac{3b}{M{-}1}$ dB SNR improvement of the sum-rate performance at high SNR (see Remark \ref{rmk:SNRgain1}). Such a sum-rate improvement is greater than the improvement achieved by single-user transmission, namely Time Division Multiple Access (TDMA), and is in contrast to ZFBF with RVQ where the sum rate saturates at high SNR. We also generalize this upper-bound to the scenario with alternating receiver-specific feedback qualities. It indicates that the sum rate performance of RS-S degrades with $\tau$ (see Remark \ref{rmk:SJMBdegrade}), where $\tau$ refers to the difference between the feedback overhead employed by the two receivers in each channel use. Moreover, we derive an upper-bound on the sum rate loss incurred by the RS-ST scheme relative to ZFBF with perfect CSIT in the scenario with alternating receiver-specific feedback qualities. It indicates that RS-ST scheme offers $3(\frac{\tau}{2(M{-}1)}{-}2)$ dB SNR gain over RS-S scheme for large value of $\tau$ (see Remark \ref{rmk:SNRgainST}).
\item To achieve a maximum allowable rate loss relative to ZFBF with perfect CSIT, equal to ${\log}_2\delta$ bps/Hz, we characterize the number of feedback bits required by the RS-S scheme in the scenarios where the two receivers have equal feedback qualities and alternating receiver-specific feedback qualities, respectively. In the former scenario, we show that compared to conventional ZFBF with RVQ, performing RS-S scheme allows for an overhead reduction that scales as $(M{-}1){\log}_2\frac{\frac{\delta}{2e}{+}\frac{e}{2}{-}1}{\sqrt{\delta}{-}1}$ at high SNR (see Remark \ref{rmk:dB1}). In the latter scenario, we show that the feedback overhead reduction offered by the RS-S scheme decreases with $\tau$ (see Remark \ref{rmk:BSJMB2}). Moreover, the number of feedback bits required by the RS-ST scheme to achieve a maximum allowable rate loss relative to ZFBF with perfect CSIT is studied in the scenario with alternating receiver-specific CSIT qualities. Compared to the RS-S scheme, performing RS-ST scheme yields a feedback overhead reduction that scales as $\frac{\tau}{2}{-}2(M{-}1)$ for large value of $\tau$.
\item Through simulation, we highlight that the RS-S and RS-ST scheme provide a significant SNR gain over the conventional (as used in LTE-A) single-user/multiuser mode switching (SU/MU) at high SNR when there is a fixed number of feedback bits.
\end{itemize}

The rest of the paper is organized as follows. Section \ref{sec:SM} elaborates on the system model and revisits RVQ, RS-S and RS-ST. The upper-bound for the sum rate loss and the feedback scaling law for RS-S scheme in the scenario with equal CSIT qualities are presented in Section \ref{sec:SJMB}. The scenario with alternating receiver-specific feedback qualities is considered in Section \ref{sec:STJMB}, where the upper-bound for the sum rate loss and feedback scaling laws for both RS-S and RS-ST schemes are studied. A performance comparison with SU/MU is shown in Section \ref{sec:performance}. Section \ref{sec:conclusion} concludes the paper.

Notations: Bold lower letters stand for vectors whereas a symbol not in bold font represents a scalar. $\left({\cdot}\right)^*$ denotes the conjugate of a scalar. $\left({\cdot}\right)^H$, $\left({\cdot}\right)^\bot$ and $\left({\cdot}\right)^\dagger$ denote the Hermitian, orthogonal space and pseudo-inverse of a matrix or vector, respectively. ${\parallel}{\cdot}{\parallel}$ is the norm of a vector. $|{\cdot}|$ is the absolute value of a complex number. $\mathbb{E}\left[{\cdot}\right]$ refers to the expectation of a random variable. $a\stackrel{d}{\sim}b$ means that random variable $a$ and $b$ are drawn from the same distribution. The notation $\angle(\mathbf{v}{,}\mathbf{w}){\triangleq}\arccos\frac{|\mathbf{v}^H\mathbf{w}|} {\parallel\mathbf{v}\parallel\parallel\mathbf{w}\parallel}$ denotes the angle between the vectors $\mathbf{v}$ and $\mathbf{w}$. 

\section{System Model}\label{sec:SM}
\begin{figure}[t]
\renewcommand{\captionfont}{\small}
\captionstyle{center}
\centering
\includegraphics[width=0.25\textwidth,height=2.25cm]{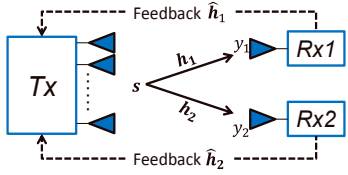}
\caption{Two-user MISO BC with quantized CSIT}\label{fig:scenario}
\end{figure}
In this paper, we consider a two-receiver MISO BC as shown in Figure \ref{fig:scenario}, where the transmitter is equipped with $M$ ($M{\geq}2$) antennas. Denoting the transmitted signal in channel use $l$ by $\mathbf{s}_l{\in}\mathbb{C}^{M{\times}1}$, subject to the power constraint ${\mathbb{E}}{\left[{\parallel}\mathbf{s}_l{\parallel}^2\right]}{\leq}P$, the received signal, $y_{kl}{\in}\mathbb{C}$ at Rx$k$ in channel use $l$, writes as
\begin{IEEEeqnarray}{rcl}
y_{kl}&{=}&\mathbf{h}_{kl}^H\mathbf{s}_l{+}\epsilon_{kl},\quad k{=}1{,}2,
\end{IEEEeqnarray}
where $\mathbf{h}_{kl}$, of size $M{\times}1$ and with $\mathcal{CN}(0{,}1)$ entries, denotes the channel vector between the transmitter and Rx$k$ in channel use $l$, and $\mathbf{h}_{kl}$ is assumed to be independent across channel uses and receivers. Here, $\epsilon_{kl}$ is the Gaussian noise with unit variance. Therefore $P$ refers to the SNR throughout the paper.

\subsection{Random Vector Quantization}
We consider a Frequency Duplex Division (FDD) setup, where the transmitter acquires the CSI through receivers' estimation and report. Since the feedback link is rate limited, vector quantization is needed and the feedback is accomplished via a finite number of bits. We assume that each receiver estimates its channel accurately and we ignore the feedback latency. Hence, the CSIT is only subject to the imperfectness due to the quantization error. In this paper, RVQ is considered as it is amenable to analysis and performs closely to optimal quantization \cite{Jin06}.

To avoid reporting the same codeword, each receiver shares a receiver-specific codebook with the transmitter. Let us employ $\mathcal{V}_{kl}{\triangleq}\{\mathbf{v}_{kl{,}1}{,}\mathbf{v}_{kl{,}2}{,}{\cdots}{,}\mathbf{v}_{kl{,}2^{B_{kl}}}\}$ to denote the codebook used by Rx$k{=}1{,}2$ in channel use $l$, where the codewords are independent and isotropically distributed in the $M$-dimensional unit sphere. The quantized CSIT is obtained as
\begin{IEEEeqnarray}{rcl}
\hat{\mathbf{h}}_{kl}&{=}&\arg\min_{\mathbf{v}_{kl{,}i}{\in}\mathcal{V}_{kl}}\sin^2\angle(\mathbf{h}_{kl}{,}\mathbf{v}_{kl{,}i}).\label{eq:hhat}
\end{IEEEeqnarray}
Afterwards, the index of the chosen codeword is quantized by Rx$k$ using $B_{kl}$ bits and reported to the transmitter. According to \cite{Jin06,Love07BasicRVQ}, the quantization error, namely $\sin^2\angle(\mathbf{h}_{kl}{,}\hat{\mathbf{h}}_{kl})$ is the minimum of $2^{B_{kl}}$ independent beta $(1{,}M{-}1)$ random variables. Its expectation is subject to
\begin{equation}
\mathbb{E}_{\mathbf{h}_{kl}{,}\mathcal{V}_{kl}}\left[\sin^2\angle(\mathbf{h}_{kl}{,}\hat{\mathbf{h}}_{kl})\right]{\leq}2^{\frac{-B_{kl}}{M{-}1}}.
\label{eq:quantizedError}
\end{equation}
Note that in this paper, only the direction of the channel is quantized, the magnitude information is not conveyed to the transmitter. This is because the magnitude information is more meaningful in performing user-selection when there is a large number of candidate users in the system \cite{Yoo07,Sharif05,Sun14THPscheduling}. As pointed out in \cite{Jin06}, when a two-receiver MISO system is considered, the magnitude feedback is of second concern.

Moreover, we consider two scenarios regarding the feedback qualities of the two receivers and the number of channel uses. The first scenario involves one channel use where the two receivers quantize their respective channels using an equal number of bits, i.e., $B_{11}{=}B_{21}{=}B$. However, in practical systems as LTE-A \cite{BrunoBook}, the feedback of CSI is receiver-specific and may only be performed on a subset of the channel uses (time and/or frequency domains). This leads to the second scenario, which consists of two channel uses and is featured by an alternating receiver-specific feedback qualities pattern. In particular, we consider that the two receivers \emph{alternatively} have a better feedback quality in the two channel uses, while they have an equal average feedback quality across the two channel uses. To be specific, the second scenario is described as $B_{11}{=}B_\beta$, $B_{21}{=}B_{\alpha}$, $B_{12}{=}B_{\alpha}$ and $B_{22}{=}B_{\beta}$, where $B_{\alpha}{<}B_{\beta}$.

\subsection{Rate-Splitting Approach}\label{sec:P1}
As it was introduced in \cite{Ges12}, in the RS-S scheme, the message intended for one receiver is split into a common and a private part, where the common part is drawn from a codebook shared by both receivers and should be decoded by both receivers with zero error probability, while the private part is to be decoded by the corresponding receiver only. The message intended for the other receiver consists of private part only. Let us use $c$ to denote the common message and $u_k$ to denote the private message intended for Rx$k$. Then, the transmitted signal in each individual channel use functions as superposing $c$ on top of ZF-precoded private messages, i.e., $u_1$ and $u_2$. Mathematically, the transmitted and received signals write as
\begin{IEEEeqnarray}{rcl}
\mathbf{s}&{=}&\underbrace{\mathbf{w}_cc}_{P_c}{+}\underbrace{\mathbf{w}_1u_1}_{P_1}{+}\underbrace{\mathbf{w}_2u_2}_{P_2},
\IEEEyessubnumber\label{eq:generalP1s}\\
y_k&{=}&\mathbf{h}_k^H\mathbf{w}_cc{+}\mathbf{h}_k^H\mathbf{w}_ku_k
{+}\mathbf{h}_k^H\mathbf{w}_ju_j{+}\epsilon_k,\, k{,}j{=}1{,}2,\, k{\neq}j,\IEEEyessubnumber
\end{IEEEeqnarray}
where the index of the channel use is ignored. The power allocation is such that $P_c{=}P(1{-}t)$ and $P_1{=}P_2{=}\frac{Pt}{2}$, where\footnote{If $t{=}0$, the common symbol is transmitted with full power and the rate is limited by the receiver with a weaker effective channel gain. This case is meaningless because it is outperformed by a single-user transmission (TDMA) whose rate is determined by the receiver with a stronger effective channel gain. Hence, we exclude this case from the support of $t$.} $t{\in}(0{,}1]$ denotes the fraction of the total power that is allocated to the private messages. Although equal power allocation for the private messages does not yield the best performance from a sum rate perspective, it allows us to find tractable results on the rate loss incurred by RS-S scheme relative to ZFBF with perfect CSIT. More details on the power optimization can be found in \cite{HamdiJMB}. The precoders are chosen as follows, for $k{=}1{,}2{,}k{\neq}j$: \footnote{Generally, the RS approach considered in this paper is a class of transmission strategies that superpose common message on top of conventional multiuser transmission. To understand the fundamental benefit of common message transmission, in most part of the paper, we consider random beamformers that improve the analytical tractability. More details on the precoder optimization can be found in \cite{HamdiJMB}.}
\begin{itemize}
\item $\mathbf{w}_k{\in}\hat{\mathbf{h}}_j^\bot$, with ${\parallel}\mathbf{w}_k{\parallel}{=}1$, is a ZF-precoder and independent of $\hat{\mathbf{h}}_k$, where $\hat{\mathbf{h}}_k$ is obtained as in \eqref{eq:hhat}.
\item $\mathbf{w}_c$, with ${\parallel}\mathbf{w}_c{\parallel}{=}1$, is a random beamformer and independent of $\mathbf{h}_k$, $\hat{\mathbf{h}}_k$ and $\mathbf{w}_k$.
\end{itemize}

\emph{Decoding:} The common message $c$ is decoded first by treating the private messages as noise. Afterwards, using SIC (i.e., removing $c$), Rx$k$ can decode $u_k$ by treating $u_j$ as noise, for $k{=}1{,}2$ and $k{\neq}j$. Consequently, the corresponding Signal-to-Interference-plus-Noise-Ratios (SINR) explicitly write as
\begin{IEEEeqnarray}{rcl}
{\rm SINR}_{c}^{(k)}&{=}&\frac{|\mathbf{h}_k^H\mathbf{w}_c|^2P(1{-}t)}{1{+}\frac{Pt}{2}\sum_{j{=}1}^2|\mathbf{h}_k^H\mathbf{w}_j|^2}
\IEEEyessubnumber\label{eq:SINRck}\\
%{\approx}\frac{|\mathbf{h}_k^H\mathbf{w}_c|^2P(1{-}t)}{1{+}|\mathbf{h}_k^H\mathbf{w}_k|^2\frac{Pt}{2}},\IEEEyessubnumber\label{eq:SINRck}\\
{\rm SINR}_c&{=}&\min({\rm SINR}_{c}^{(1)}{,}{\rm SINR}_{c}^{(2)}),\IEEEyessubnumber\label{eq:SINRc}\\
{\rm SINR}_k&{=}&\frac{|\mathbf{h}_k^H\mathbf{w}_k|^2\frac{Pt}{2}}{1{+}|\mathbf{h}_k^H\mathbf{w}_j|^2\frac{Pt}{2}},k{\neq}j.
\IEEEyessubnumber\label{eq:SINRk}
\end{IEEEeqnarray}
The ergodic rate of each message is expressed as a function of the power splitting ratio, namely $R_c(t){\triangleq}\mathbb{E}\left[{\log}_2(1{+}{\rm SINR}_c)\right]$ and $R_k(t){\triangleq}\mathbb{E}\left[{\log}_2(1{+}{\rm SINR}_k)\right]$. \begin{myremark}\label{rmk:Rcsplit}
In a more general RS approach, the messages of both receivers are split into a common part, $m_{ck}$, and a private part, $m_{pk}$, for $k{=}1{,}2$. The private parts, $m_{p1}$ and $m_{p2}$, are transmitted similarly to $u_1$ and $u_2$, while $c$ is a general common message which can be a mixture of $m_{c1}$ and $m_{c2}$. Then, any non-negative rates $R_{c1}$ and $R_{c2}$ such that $R_{c1}{+}R_{c2}{=}R_c$ are achievable by properly splitting the bits encoded in $c$. Hence, assuming $c$ is made up of either $m_{c1}$ or $m_{c2}$ is a special case. As we focus on a sum rate analysis, it suffices to consider rate splitting for only one receiver.
\end{myremark}

\underline{\emph{Connection with \cite{Ges12,pimrc2013,Elia13,IT_MISO_BC}}}: We point out that the RS-S scheme proposed in \cite{Ges12,pimrc2013,Elia13,IT_MISO_BC} (Lemma 2 in \cite{Ges12}, ``matched case'' in \cite{pimrc2013}, Scheme $\mathcal{X}_3$ in \cite{Elia13} and $\mathcal{P}_1$, $\mathcal{Q}_1$ scheme in \cite{IT_MISO_BC}) is investigated from a DoF perspective. When the two receivers have equal feedback qualities, if $B$ scales with SNR as $B{=}\alpha(M{-}1){\log}_2P{+}o({\log}_2P)$ (where $0{\leq}\alpha{\leq}1$), using \eqref{eq:quantizedError}, one can easily obtain that the quantization error decays as $P^{-\alpha}$. According to the findings in \cite{Ges12,pimrc2013,Elia13,IT_MISO_BC}, by choosing $P_c{=}P{-}P^{\alpha}$ and $P_1{=}P_2{=}\frac{P^{\alpha}}{2}$, the residual interference $|\mathbf{h}_k^H\mathbf{w}_j|^2\frac{Pt}{2}{,}k{\neq}j$ will be received with a power similar to the noise. Then, RS-S scheme achieves the sum DoF of $1{+}\alpha$, which is greater than $2\alpha$ that is achieved by ZFBF, i.e., with $P_c{=}0$ and $P_1{=}P_2{=}\frac{P}{2}$. Note that, although RS-S scheme has no DoF gain over TDMA for $\alpha{=}0$ and ZFBF with RVQ for $\alpha{=}1$, its benefits over TDMA and ZFBF with RVQ in terms of sum rate and feedback overhead requirement remain to be investigated. This is the main focus of Section \ref{sec:SJMB}.

In the scenario with alternating receiver-specific feedback qualities, if $B_\alpha{=}\alpha(M{-}1){\log}_2P{+}o({\log}_2P)$ and $B_\beta{=}\beta(M{-}1){\log}_2P{+}o({\log}_2P)$ where $0{\leq}\alpha{<}\beta{\leq}1$, \cite{IT_MISO_BC,Jinyuan_evolving_misobc} suggested that performing RS-S scheme in each channel use with the power allocation $P_1{=}P_2{=}\frac{P^\alpha}{2}$ yields the sum DoF $1{+}\alpha$. However, such a result does not reveal the usefulness of having alternating CSIT qualities, i.e., $B_{11}{>}B_{21}$ and $B_{12}{<}B_{22}$. This leads to the emergence of a space-time design of the RS approach (RS-ST).

\subsection{Rate-Splitting Approach with Space-Time design}\label{sec:P2}
In the scenario with alternating receiver-specific feedback qualities, the RS-ST scheme was proposed in \cite{Tandon12,Elia13} (Scheme $S_3^{3/2}$ in \cite{Tandon12} and Scheme $\mathcal{X}_2$ in \cite{Elia13}) to enhance the sum DoF achieved with the RS-S scheme. Comparing with the RS-S scheme, RS-ST scheme transmits an additional common message (resulted by a further split of the messages), i.e., $c_0$, across the two channel uses. Specifically, the transmitted signals in channel use 1 and 2 write as
\begin{IEEEeqnarray}{rcl}
\mathbf{s}_1&{=}&\underbrace{\mathbf{w}_{c1}c_1}_{P(1{-}t_\beta)}{+}\underbrace{\mathbf{w}_{01}c_0}_{P(t_\beta{-}t_\alpha)/2}{+}
\underbrace{\mathbf{w}_{11}u_{11}}_{Pt_\alpha/2}{+}\underbrace{\mathbf{w}_{21}u_{21}}_{Pt_\beta/2},\IEEEyessubnumber\label{eq:s1STJMB}\\
\mathbf{s}_2&{=}&\underbrace{\mathbf{w}_{c2}c_2}_{P(1{-}t_\beta)}{+}\underbrace{\mathbf{w}_{02}c_0}_{P(t_\beta{-}t_\alpha)/2}{+}
\underbrace{\mathbf{w}_{12}u_{12}}_{Pt_\beta/2}{+}\underbrace{\mathbf{w}_{22}u_{22}}_{Pt_\alpha/2},\IEEEyessubnumber\label{eq:s2STJMB}
\end{IEEEeqnarray}
respectively, where $u_{kl}$ denotes the symbol that carries the private message intended for Rx$k$ in channel use $l$, $c_l$ is the common messages transmitted in channel use $l$. The power of $c_0$ is chosen as the difference between the powers allocated to the private messages in each channel use, namely $P\frac{t_{\beta}{-}t_\alpha}{2}$. $t_\beta$ and $t_\alpha$ are the power splitting ratios, where $0{<}t_{\alpha}{\leq}t_{\beta}{\leq}1$. The precoders are chosen as follows:
\begin{itemize}
\item $\mathbf{w}_{cl}$ and $\mathbf{w}_{kl}$, $k{=}1{,}2$, are respectively the random beamformer and ZF-precoders in channel use $l$, similar to the RS-S scheme;
\item We choose $\mathbf{w}_{01}{=}\mathbf{w}_{11}{\in}\hat{\mathbf{h}}_{21}^\bot$ and $\mathbf{w}_{02}{=}\mathbf{w}_{22}{\in}\hat{\mathbf{h}}_{12}^\bot$. Although such choice is non-optimal, it suffices to provide the fundamental benefit of transmitting $c_0$ across the two channel uses.
\end{itemize}
This leads to the following received signals for $k{=}1{,}2$ and $j{\neq}k$,
\begin{IEEEeqnarray}{rcl}
y_{k1}&{=}&\mathbf{h}_{k1}^H\mathbf{w}_{c1}c_1{+}\mathbf{h}_{k1}^H\mathbf{w}_{01}c_0{+}\nonumber\\
&&\mathbf{h}_{k1}^H\mathbf{w}_{k1}u_{k1}{+}\mathbf{h}_{k1}^H\mathbf{w}_{j1}u_{j1}{+}\epsilon_{k1},\IEEEyessubnumber\label{eq:yk1}\\
y_{k2}&{=}&\mathbf{h}_{k2}^H\mathbf{w}_{c2}c_2{+}\mathbf{h}_{k2}^H\mathbf{w}_{02}c_0{+}\nonumber\\
&&\mathbf{h}_{k2}^H\mathbf{w}_{k2}u_{k2}{+}\mathbf{h}_{k2}^H\mathbf{w}_{j2}u_{j2}{+}\epsilon_{k2}.\IEEEyessubnumber\label{eq:yk2}
\end{IEEEeqnarray}

\emph{Decoding}: Let us focus on the decoding at Rx1. Following the decoding process elaborated in \cite{Tandon12,Elia13}, using SIC, Rx1 firstly decodes $c_1$ and $c_0$ sequentially in $y_{11}$ by treating the private messages as noise. Secondly, after removing $c_0$ from $y_{12}$, Rx1 recovers $c_2$ by treating the private messages as noise. Thirdly, by removing all the common messages, Rx1 decodes $u_{11}$ and $u_{12}$ in channel use 1 and 2 respectively. Similarly, Rx2 decodes $c_2$ and $c_0$ from $y_{22}$, recovers $c_1$ from $y_{21}$ and proceeds to decode the private messages afterwards. The SINR of the messages decoded by Rx1 are explicitly written as
\begin{IEEEeqnarray}{rcl}\label{eq:SINRst}
\!\!\!\!&&{\rm SINR}_{c1}^{(1)}{=}\nonumber\\
\!\!\!\!&&\frac{|\mathbf{h}_{11}^H\mathbf{w}_{c1}|^2P(1{-}t_\beta)}
{1{+}|\mathbf{h}_{11}^H\mathbf{w}_{01}|^2\frac{P(t_\beta{-}t_\alpha)}{2}{+}
|\mathbf{h}_{11}^H\mathbf{w}_{11}|^2\frac{Pt_\alpha}{2}{+}|\mathbf{h}_{11}^H\mathbf{w}_{21}|^2\frac{Pt_\beta}{2}},
\IEEEyessubnumber\label{eq:c1k}\\
\!\!\!\!&&{\rm SINR}_{c2}^{(1)}{=}\frac{|\mathbf{h}_{12}^H\mathbf{w}_{c2}|^2P(1{-}t_\beta)}
{1{+}|\mathbf{h}_{12}^H\mathbf{w}_{12}|^2\frac{Pt_\beta}{2}{+}|\mathbf{h}_{12}^H\mathbf{w}_{22}|^2\frac{Pt_\alpha}{2}},%\nonumber\\
\IEEEyessubnumber\label{eq:c2k}\\
\!\!\!\!&&{\rm SINR}_{c0}^{(1)}{=}\frac{|\mathbf{h}_{11}^H\mathbf{w}_{01}|^2\frac{P(t_\beta{-}t_\alpha)}{2}}
{1{+}|\mathbf{h}_{11}^H\mathbf{w}_{11}|^2\frac{Pt_\alpha}{2}{+}|\mathbf{h}_{11}^H\mathbf{w}_{21}|^2\frac{Pt_\beta}{2}},%\nonumber\\
\IEEEyessubnumber\label{eq:c0k}\\
\!\!\!\!&&{\rm SINR}_{11}{=}\frac{|\mathbf{h}_{11}^H\mathbf{w}_{11}|^2\frac{Pt_\alpha}{2}}
{1{+}|\mathbf{h}_{11}^H\mathbf{w}_{21}|^2\frac{Pt_\beta}{2}},\IEEEyessubnumber\label{eq:ukeql}\\
\!\!\!\!&&{\rm SINR}_{12}{=}\frac{|\mathbf{h}_{12}^H\mathbf{w}_{12}|^2\frac{Pt_\beta}{2}}
{1{+}|\mathbf{h}_{12}^H\mathbf{w}_{22}|^2\frac{Pt_\alpha}{2}}.\IEEEyessubnumber\label{eq:ukneql}
\end{IEEEeqnarray}
The SINR of the messages decoded by Rx2 are omitted for brevity as they write similarly. The ergodic rate is computed by $R_{cl}(t_\beta{,}t_\alpha){\triangleq}\mathbb{E}[{\log}_2(1{+}\min_{k{=}1{,}2}{\rm SINR}_{cl}^{(k)})]$, $l{=}0{,}1{,}2$ and $R_{kl}(t_\beta{,}t_\alpha){\triangleq}\mathbb{E}\left[{\log}_2(1{+}{\rm SINR}_{kl})\right]$.

\underline{\emph{Connection with \cite{Tandon12,Elia13}}}: We point out that \cite{Tandon12,Elia13} show the benefit of the space-time transmission of $c_0$ from a DoF perspective. Considering $B_\alpha{=}\alpha(M{-}1){\log}_2P{+}o({\log}_2P)$ and $B_\beta{=}\beta(M{-}1){\log}_2P{+}o({\log}_2P)$ where $0{\leq}\alpha{<}\beta{\leq}1$, the quantization errors incurred in $\hat{\mathbf{h}}_{12}$ and $\hat{\mathbf{h}}_{21}$ decay as $P^{-\alpha}$ and the quantization errors incurred in $\hat{\mathbf{h}}_{11}$ and $\hat{\mathbf{h}}_{22}$ decay as $P^{-\beta}$. According to \cite{Tandon12,Elia13}, with $Pt_\alpha{=}P^\alpha$ and $Pt_\beta{=}P^\beta$, the residual interference $|\mathbf{h}_{12}^H\mathbf{w}_{22}|^2\frac{Pt_\alpha}{2}$ and $|\mathbf{h}_{11}^H\mathbf{w}_{21}|^2\frac{Pt_\beta}{2}$ will be received with a power similar to the noise. The sum DoF achieved by $c_1$, $c_2$ and all the private messages is $1{+}\alpha$, while $c_0$ achieves the DoF of $\frac{\beta{-}\alpha}{2}$. Thus, the resultant sum DoF is $1{+}\frac{\alpha{+}\beta}{2}$, which is greater than $1{+}\alpha$ that is achieved with the RS-S scheme. However, the benefits of the space-time transmission over RS-S scheme in terms of sum rate and feedback overhead requirement remains to be investigated. This is the main focus in Section \ref{sec:STJMB}.

Next, we will carry out some preliminary calculations for the random variables involved in the SINR expression, followed by the analysis on the sum rate and the feedback overhead reduction. 

\section{RS-S with equal feedback qualities}\label{sec:SJMB}
In this section, we focus on the scenario where the two receivers have equal feedback qualities, i.e., $B_{11}{=}B_{21}{=}B$. Before going into the main results, some preliminary results are derived as they are frequently used in the rest of the paper.
\subsection{Preliminary Calculations}
\begin{figure}[t]
\renewcommand{\captionfont}{\small}
\captionstyle{center}
\centering
\subfigure[Joint CDF vs. $M$]{%{0.32\textwidth}
                \centering
                \includegraphics[width=0.23\textwidth,height=3.5cm]{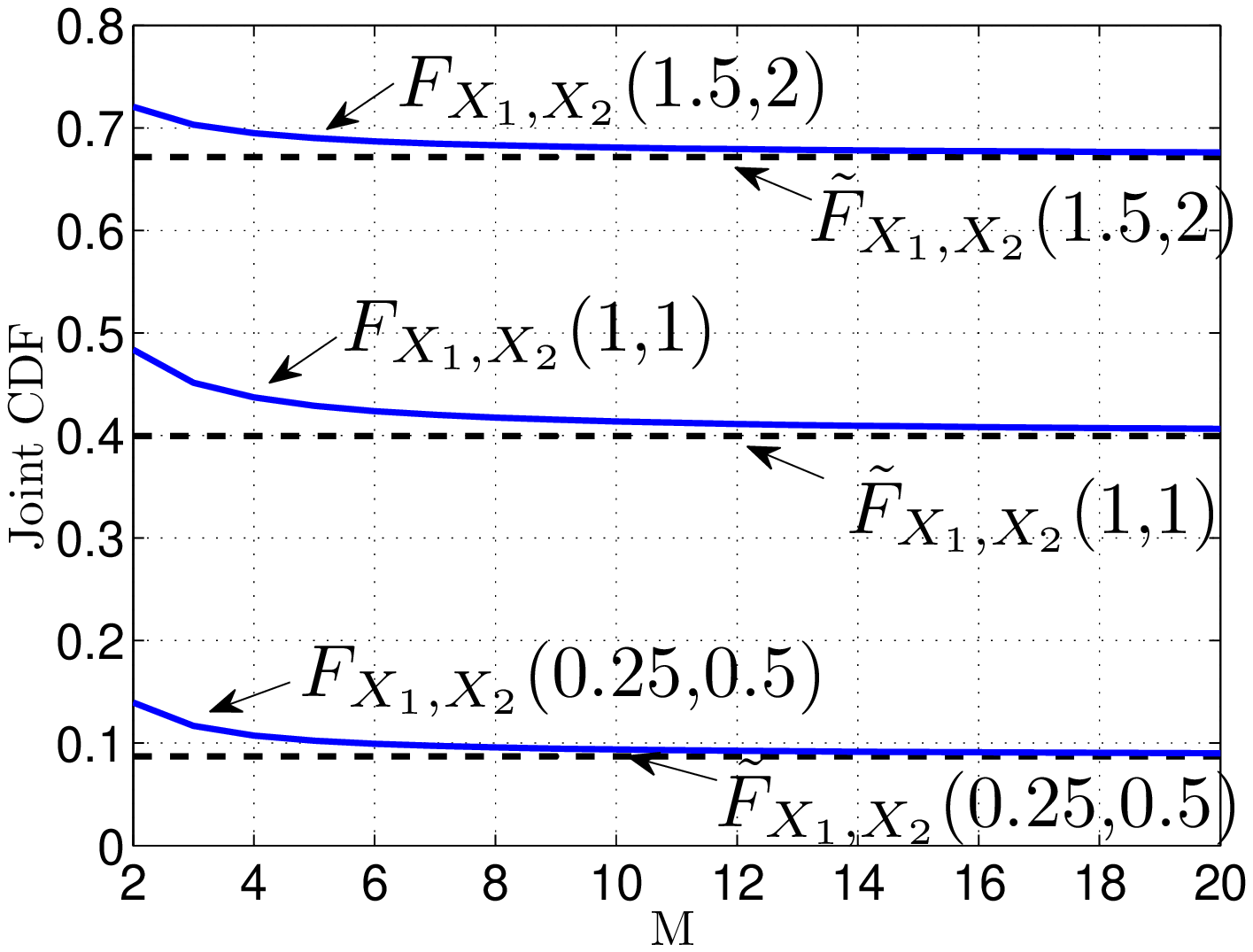}
                %\caption{$\mathcal{D}_1$}
                \label{fig:JCDFvsM}
        }%\end{subfigure}
\subfigure[CDF of $Y_k$, 30dB, $t{=}0.2$ and $B{=}10$]{%{0.32\textwidth}
                \centering
                \includegraphics[width=0.26\textwidth,height=3.5cm]{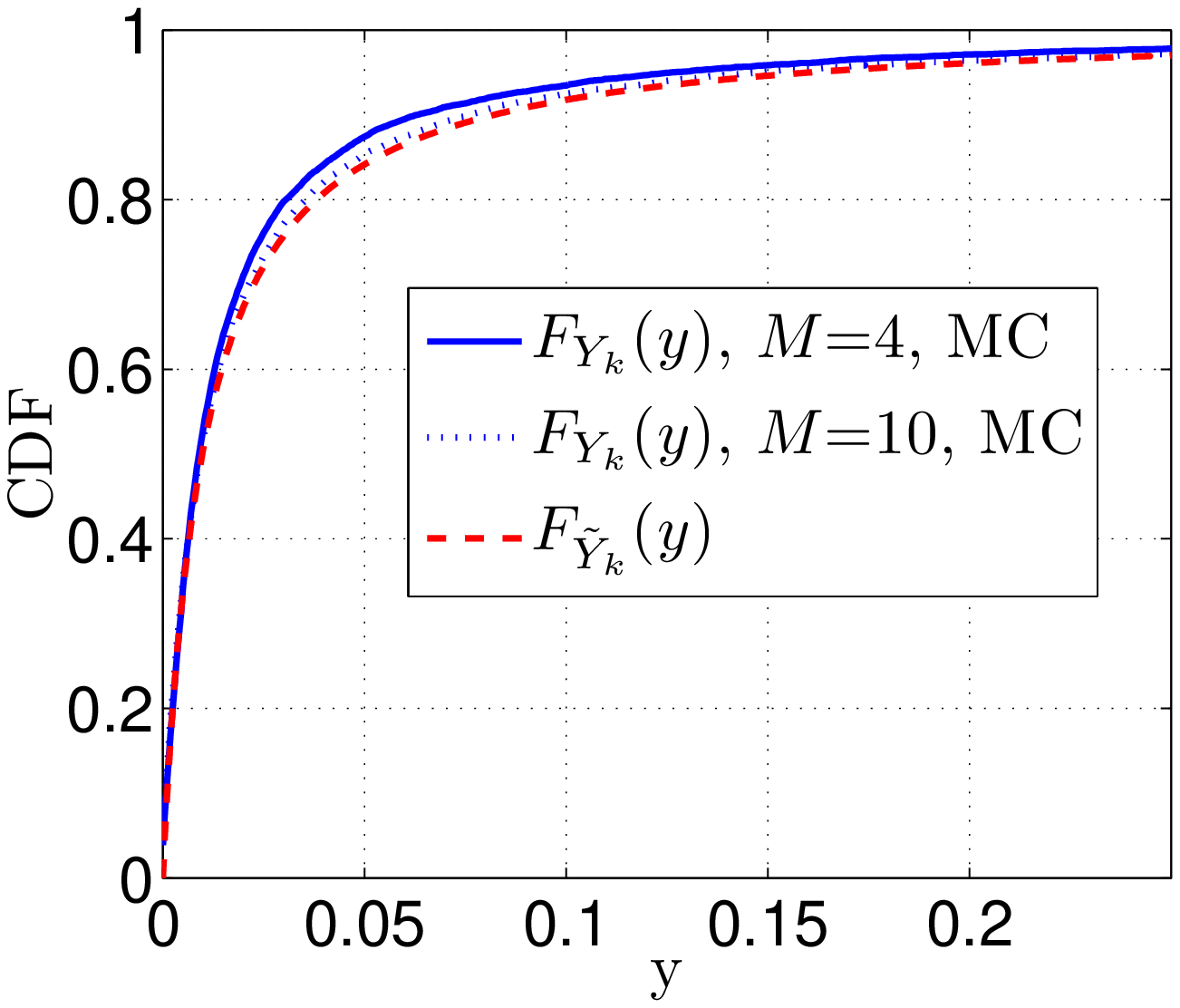}
                %\caption{$\mathcal{D}_1$}
                \label{fig:CDFyk}
        }%\end{subfigure}
\caption{CDF approximation}\label{fig:cdfapprox}
\end{figure}

\begin{mylemma}\label{lemma:upperQuantizeError}
\cite[Lemma 1 and 2]{Jin06} The random variable, $|\bar{\mathbf{h}}_k^H\mathbf{w}_j|^2{,}k{\neq}j$, where $\bar{\mathbf{h}}_k{=}\frac{\mathbf{h}_k}{{\parallel}\mathbf{h}_k{\parallel}}$, is equal to the product of the quantization error ${\angle}{\sin}^2(\mathbf{h}_k{,}\hat{\mathbf{h}}_k)$ and a beta $(1{,}M{-}2)$ random variable. Note that ${\angle}{\sin}^2(\mathbf{h}_k{,}\hat{\mathbf{h}}_k)$ and the beta $(1{,}M{-}2)$ random variable are independent of each other. The expectation of $|\bar{\mathbf{h}}_k^H\mathbf{w}_j|^2{,}k{\neq}j$ is subject to
\begin{equation}
\mathbb{E}_{\mathbf{h}_k{,}\mathcal{V}_k}\left[|\bar{\mathbf{h}}_k^H\mathbf{w}_j|^2\right]{<}\frac{1}{M{-}1}2^{\frac{-B_k}{M{-}1}},k{\neq}j.
\end{equation}
\end{mylemma}
As it will be seen in the proof of Proposition \ref{theo:RateLossP1}, \ref{prop:ratelossQ1} and \ref{prop:RateLossP2}, Lemma \ref{lemma:upperQuantizeError} is used to upper-bound the rate loss incurred by the ZF precoders in the RS-S (and RS-ST) scheme with RVQ.

Next, we aim to compute the distribution of ${\rm SINR}_c^{(k)}$ and the distribution of the minimum of ${\rm SINR}_c^{(1)}$ and ${\rm SINR}_c^{(2)}$. Towards this, we introduce the following assumption to ease the computation complexity.
\begin{myasump}\label{asp:sinr_approx}
We assume that the feedback qualities are good enough for both receivers, such that the $|\mathbf{h}_k^H\mathbf{w}_k|^2{\gg}|\mathbf{h}_k^H\mathbf{w}_j|^2$ in \eqref{eq:SINRck} and $|\mathbf{h}_{kl}^H\mathbf{w}_{kl}|^2{\gg}|\mathbf{h}_{kl}^H\mathbf{w}_{jl}|^2$ in \eqref{eq:c1k}, \eqref{eq:c2k} and \eqref{eq:c0k} hold with a high likelihood, where $k{\neq}j$. Then, by introducing $X_{k1}{\triangleq}|\mathbf{h}_k^H\mathbf{w}_c|^2$, $X_{k2}{\triangleq}|\mathbf{h}_k^H\mathbf{w}_k|^2$ and $Y_k{\triangleq}\frac{X_{k1}}{1{+}X_{k2}\frac{Pt}{2}}$, ${\rm SINR}_c^{(k)}$ in \eqref{eq:SINRck} and ${\rm SINR}_c$ in \eqref{eq:SINRc} are approximated by $P(1{-}t)Y_k$ and $P(1{-}t)Y$ with $Y{\triangleq}{\min}(Y_1{,}Y_2)$, respectively. The approximations of \eqref{eq:c1k}, \eqref{eq:c2k} and \eqref{eq:c0k} follow similarly.
\end{myasump}

Note that Assumption \ref{asp:sinr_approx} is only applied to the derivation of the rate of the common messages. The impact of the residual interference after ZFBF with RVQ is considered in the derivation of the rate of the private messages. However, in the simulation of Section \ref{sec:SJMBrate}, \ref{sec:SJMBB}, \ref{sec:STJMB} and \ref{sec:performance}, the SINR of the common messages are calculated following \eqref{eq:SINRck}, \eqref{eq:c1k}, \eqref{eq:c2k} and \eqref{eq:c0k}.% As we will see later on, although this assumption upper-bounds the SINR (i.e., lower-bounds sum rate loss relative to ZFBF with perfect CSIT), simulation results verify that 1) Proposition \ref{theo:RateLossP1}, \ref{prop:ratelossQ1} and \ref{prop:RateLossP2} still upper-bound the sum rate loss, and 2) the maximum allowable rate loss is satisfied given the feedback overhead stated in Proposition \ref{theo:B} and \ref{coro:BSrs} and eq. \eqref{eq:BST}.}

To calculate the distribution of ${\rm SINR}_c$, it suffices to study the distribution of $Y$. To this end, we calculate the joint distribution of $X_{k1}$ and $X_{k2}$. We observe that $X_{k1}$ and $X_{k2}$ are exponential distributed with parameter $1$, because $\mathbf{h}_k$ is a complex Gaussian vector and $\mathbf{w}_c$ and $\mathbf{w}_k$ are isotropic unit vectors independent of $\mathbf{h}_k$. Moreover, we see that $X_{k1}$ and $X_{k2}$ are correlated as both of them depend on the realization of $\mathbf{h}_k$. Their joint distribution is characterized as follows.
\begin{mylemma}\label{lemma:jcdf}
The joint cumulative distribution function (CDF) of the correlated exponential random variables $X_{k1}{=}|\mathbf{h}_k^H\mathbf{w}_c|^2$ and $X_{k2}{=}|\mathbf{h}_k^H\mathbf{w}_k|^2$ is given by
\begin{equation}
F_{X_{k1}{,}X_{k2}}(x_1{,}x_2){=}1{-}e^{-x_1}{-}e^{-x_2}{+}\xi(x_1{,}x_2),\label{eq:jcdf}
\end{equation}
\begin{figure*}[t]
\begin{equation}
\xi(x_1{,}x_2){=}\frac{1}{\Gamma(M)}\sum_{i{=}0}^{M{-}1}\sum_{j{=}0}^{M{-}1}
(-x_1)^{M{-}1{-}i}(-x_2)^{M{-}1{-}j}{{M{-}1}\choose{i}}{{M{-}1}\choose{j}}\Gamma(i{+}j{+}2{-}M{,}\max(x_1{,}x_2)),\label{eq:xi}
\end{equation}
\hrulefill
\end{figure*}where $x_1{,}x_2{\in}[0{,}\infty)$ and $\xi(x_1{,}x_2)$ is given in \eqref{eq:xi} at the top of next page. $\Gamma(r){=}(r{-}1)!$ is the Gamma function for positive integer $r$, while $\Gamma(r{,}a){=}\int_a^{\infty}a^re^{-a}da$ refers to the Upper Incomplete Gamma function, which is also valid for $r{\leq}0$.
\end{mylemma}
\emph{Proof:} see Appendix A.$\hfill\Box$

Note that it is cumbersome to utilize \eqref{eq:jcdf} to obtain the distribution of $Y$ and perform analysis. Hence, we approximate \eqref{eq:jcdf} by assuming that $X_{k1}$ and $X_{k2}$ are independent, namely
\begin{IEEEeqnarray}{rcl}
F_{X_{k1}{,}X_{k2}}(x_1{,}x_2)&{\approx}&F_{\tilde{X}_{k1}{,}\tilde{X}_{k2}}(x_1{,}x_2)\nonumber\\
&{=}&1{-}e^{-x_1}{-}e^{-x_2}{+}e^{-x_1-x_2},\label{eq:jcdfapprox}
\end{IEEEeqnarray}
where $\tilde{X}_{k1}\stackrel{d}{\sim}X_{k1}$, $\tilde{X}_{k2}\stackrel{d}{\sim}X_{k2}$ and $\tilde{X}_{k1}$ and $\tilde{X}_{k2}$ are independent. Figure \ref{fig:JCDFvsM} shows that the approximation is good for sufficiently large value of $M$, and it is good enough for $M{=}4$. Hence, we employ \eqref{eq:jcdfapprox} instead of \eqref{eq:jcdf} in the subsequent derivations to make the analysis more tractable.

Let us introduce $\tilde{Y}_k{\triangleq}\frac{\tilde{X}_{k1}}{1{+}\frac{Pt}{2}\tilde{X}_{k2}}$, which is an approximation of $Y_k{\triangleq}\frac{X_{k1}}{1{+}X_{k2}\frac{Pt}{2}}$ since $F_{X_{k1}{}{,}X_{k2}}(x_1{,}x_2){\approx}F_{\tilde{X}_{k1}{,}\tilde{X}_{k2}}(x_1{,}x_2)$ in \eqref{eq:jcdfapprox}. Since $\tilde{X}_{k1}$ and $\tilde{X}_{k2}$ are independent, we compute the CDF of $\tilde{Y}_k$ as
\begin{IEEEeqnarray}{rcl}
F_{\tilde{Y}_k}(y)&{=}&\int_0^{\infty}{\rm Pr}(\tilde{X}_{k1}{<}y(1{+}\frac{Pt}{2}x))f_{\tilde{X}_{k2}}(x)dx\nonumber\\ 
&{=}&1{-}\frac{e^{-y}}{1+\frac{Pt}{2}y}{\approx}F_{Y_k}(y),\label{eq:cdfyk}
\end{IEEEeqnarray}
using the fact that $\tilde{X}_{k1}$ and $\tilde{X}_{k2}$ are exponential distributed with parameter $1$. A comparison of $F_{Y_k}$ and $F_{\tilde{Y}_k}$ is shown by Figure \ref{fig:CDFyk}, where $F_{Y_k}(y)$ is plotted via Monte Carlo (MC) simulation.

Next, we study the distribution of $Y{=}{\min}(Y_1{,}Y_2)$. As $\mathbf{w}_k$ is isotropically chosen from the null space of $\hat{\mathbf{h}}_j{,}j{\neq}k$ and $\hat{\mathbf{h}}_j$ is obtained using \eqref{eq:hhat}, we can see that $\mathbf{w}_k$ is correlated with $\mathbf{h}_j{,}j{\neq}k$. In turn, it follows that $Y_1$ and $Y_2$ are correlated. Thus, it is cumbersome to derive the exact distribution of $Y{=}{\min}(Y_1{,}Y_2)$. Instead, we provide an upper-bound of the CDF of $Y$ as follows.
\begin{mylemma}\label{lemma:CDFUB}
(Upper-bound on the CDF of $Y$)
\begin{IEEEeqnarray}{rcl}
F_{Y}(y)&{\leq}&1{-}(1{-}F_{Y_1}(y))^2.
\end{IEEEeqnarray}
\end{mylemma}
\emph{Proof:} The inequality directly follows \cite[eq (5.4.1b)]{Order_Statistic}. $\hfill\Box$

Using \eqref{eq:cdfyk}, an approximation of this upper-bound writes as
\begin{IEEEeqnarray}{rcl}
\!\!\!\!F_{Y}(y)&{\leq}&1{-}(1{-}F_{Y_1}(y))^2\nonumber\\
\!\!\!\!&{\approx}&1{-}(1{-}F_{\tilde{Y}_1}(y))^2{=}1{-}\frac{1}{(1{+}\frac{Pt}{2}y)^2}e^{-2y},y{\in}[0{,}\infty).\label{eq:cdfy}
\end{IEEEeqnarray}

Moreover, we introduce the following useful Lemma.
\begin{mylemma}\label{lemma:ElogLB}
For random variables $Z$ and $\tilde{Z}$ who have the same support $(-\infty{,}\infty)$ and whose CDF satisfy $F_Z(z){\leq}F_{\tilde{Z}}(z)$, we have $\mathbb{E}[Z]{\geq}\mathbb{E}[\tilde{Z}]$.
\end{mylemma}
\emph{Proof:} see Appendix B. $\hfill\Box$

In the proof of Proposition \ref{theo:RateLossP1}, \ref{prop:ratelossQ1} and \ref{prop:RateLossP2}, the rate of the common message is lower-bounded by a function of $\mathbb{E}[{\ln}Y]$. Lemma \ref{lemma:ElogLB} allows us to lower-bound $\mathbb{E}[{\ln}Y]$ using the right hand side (r.h.s.) of \eqref{eq:cdfy}.

Next, we study the sum rate loss incurred by the RS-S scheme relative to the ZFBF with perfect CSIT and investigate the scaling law of $B$ to achieve a maximum allowable rate loss.
\subsection{Sum Rate Loss}\label{sec:SJMBrate}
\begin{figure}[t]
\renewcommand{\captionfont}{\small}
\captionstyle{center}
\centering
\subfigure[Sum rate loss, $B{=}10$]{%{0.32\textwidth}
                \centering
                \includegraphics[width=0.25\textwidth,height=3.5cm]{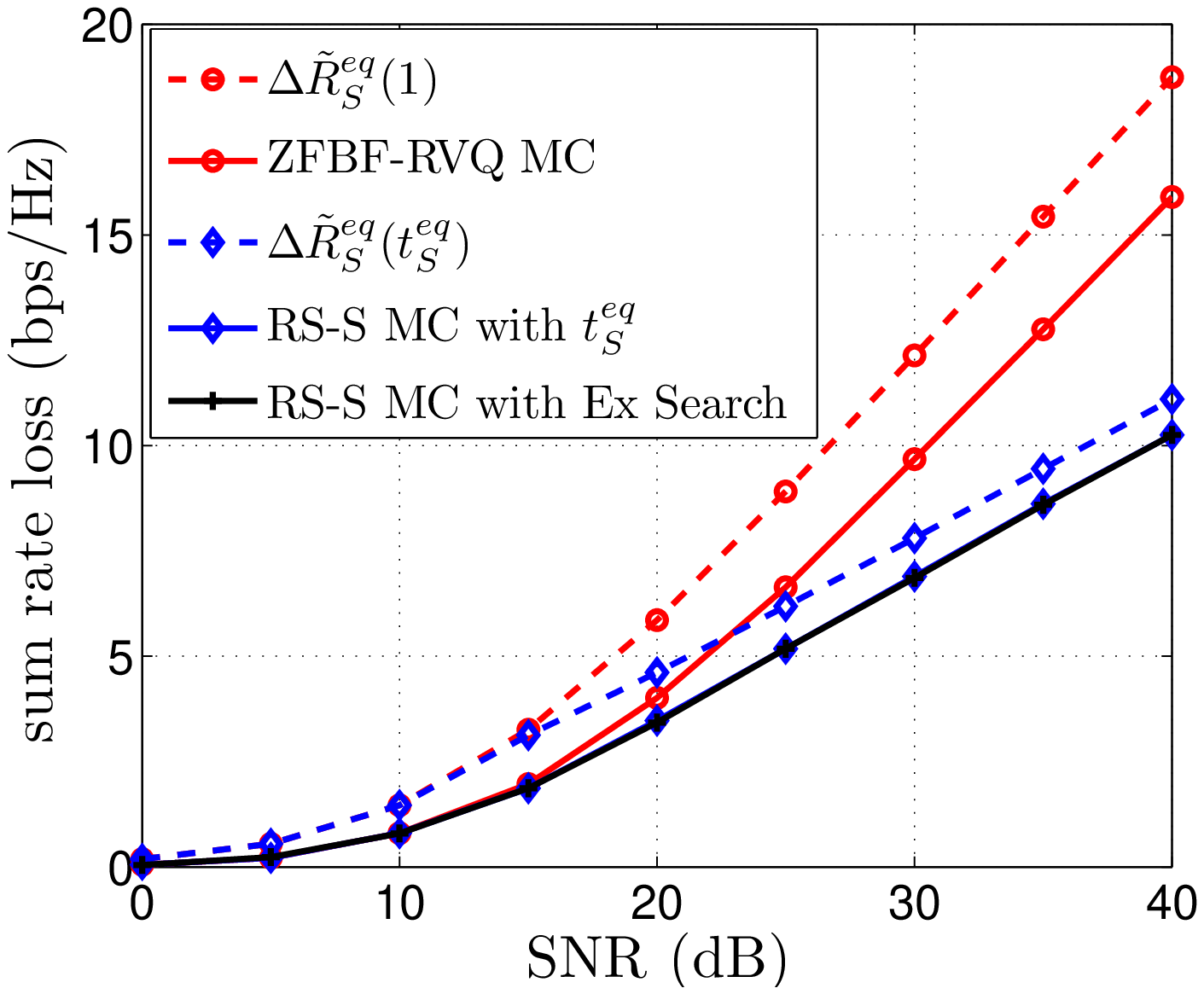}
                %\caption{$\mathcal{D}_2$}
                \label{fig:RLM4B10}
        }%\end{subfigure}
\subfigure[Sum rate]{%{0.32\textwidth}
                \centering
                \includegraphics[width=0.24\textwidth,height=3.5cm]{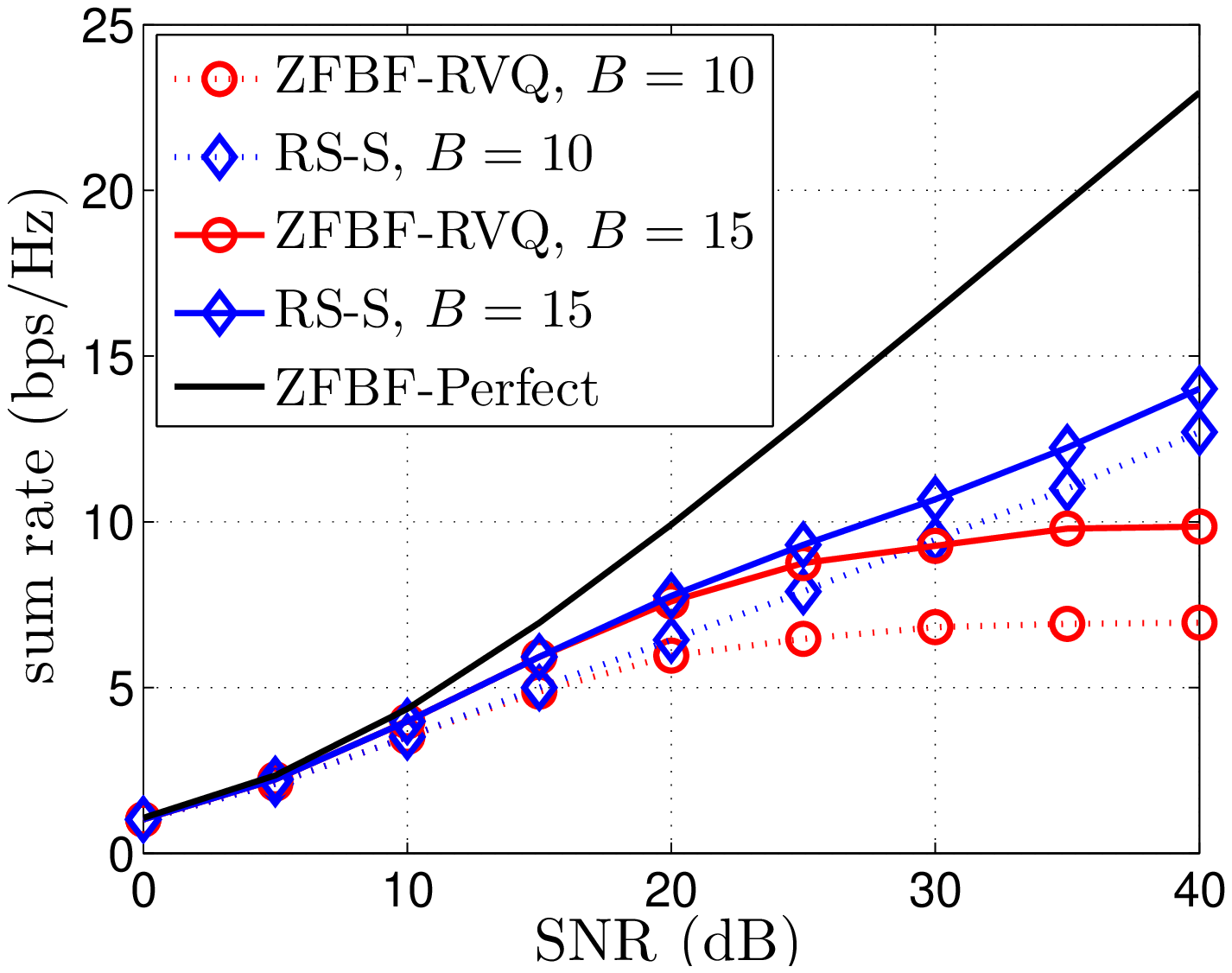}
                %\caption{$\mathcal{D}_1$}
                \label{fig:RsM4B10}
        }%\end{subfigure}
\caption{Simulation results for RS-S with $M{=}4$.}\label{fig:naiveP1}
\end{figure}
To study the sum rate loss incurred by the RS-S scheme in the scenario where the two receivers have equal feedback qualities, we define $\Delta R_S^{eq}(t){\triangleq}R_1^p{+}R_2^p{-}R_1(t){-}R_2(t){-}R_c(t)$ to be the difference between the sum rate achieved with ZFBF with perfect CSIT and the sum rate achieved with RS-S with a power splitting ratio $t{\in}(0{,}1]$. The expression $R_k^p{\triangleq}\mathbb{E}\left[{\log_2}(1{+}|\mathbf{h}_k^H\mathbf{w}_{k{,}pf}|^2\frac{P}{2})\right]$ denotes the rate achieved by Rx$k$, $k{=}1{,}2$, using ZFBF with perfect CSIT, where $\mathbf{w}_{k{,}pf}$ is a unit-norm vector randomly chosen from the $M{-}1$-dimensional null space of $\mathbf{h}_j{,}k{\neq}j$. An upper-bound of $\Delta R_S^{eq}(t)$ is stated below.

\begin{myprop}\label{theo:RateLossP1}
In the scenario where the two receivers have equal feedback qualities, the sum-rate loss incurred by the RS-S scheme with RVQ relative to the ZFBF with perfect CSIT is upper-bounded by
\begin{multline}
\Delta R_{S}^{eq}(t){\leq}\Delta \tilde{R}_S^{eq}(t){=}2\epsilon(t){+}
2{\log}_2(1{+}\frac{Pt M}{2(M{-}1)}2^{\frac{-B}{M-1}}){-}\\{\log}_2(1{+}\frac{P(1{-}t)}{2}e^{\kappa(t)}),\label{eq:dRt}
\end{multline}
where $\kappa(t){\triangleq}(\frac{4}{Pt}{-}1)\phi(\frac{Pt}{4}){-}1{-}\gamma$, $\epsilon(t){\triangleq}\frac{1}{\ln2}\left[\phi(\frac{P}{2}){-}\phi(\frac{Pt}{2})\right]$, while $t{\in}(0{,}1]$ is the power splitting ratio, $\gamma{\approx}0{.}577$ is the Euler constant, $\phi(x){\triangleq}e^{\frac{1}{x}}E_1(\frac{1}{x})$ with $E_1(x){=}\int_1^\infty\frac{e^{-xt}}{t}dt$ and $e{\approx}2.718$ refers to the natural constant.
\end{myprop}
\emph{Proof:} see Appendix C. $\hfill\Box$

In \eqref{eq:dRt}, the first term $2\epsilon(t)$ stands for the rate loss due to the decrement of the power allocated to the private messages, the second term which is a function of $B$ refers to the rate loss incurred by the ZF precoders (with power $Pt$) of RS-S with RVQ, while the last term is the rate achieved by the common message, i.e., $c$. Taking $t{=}1$ yields the rate loss incurred by the conventional ZFBF with RVQ, where the first and last term become zero.\footnote{We note that the expression of the upper-bound of the sum rate loss incurred by ZFBF with RVQ is different from that is derived in \cite{Jin06} due to the following reasons: 1) we consider a $M{\times}2$ system while \cite{Jin06} studied a $M{\times}M$ system, and 2) we consider that the ZF precoder is randomly chosen from the null space of the unintended receiver, whereas \cite{Jin06} obtained the ZF precoders by computing the pseudo-inverse of the aggregate channel.}

Next, we aim to find the closed-form solution of the optimal power splitting ratio $t^*{\triangleq}\arg\min_{0{<}t{\leq}1} \Delta\tilde{R}_S^{eq}(t)$, but it is difficult to obtain for arbitrary SNR due to the complicated expression of $\Delta\tilde{R}_S^{eq}(t)$. Hence, to improve the analysis tractability, we consider high SNR regime and aim to obtain $t_S^{eq}{\triangleq}\arg\min_{0{<}t{\leq}1} \Delta\tilde{R}_S^{eq}(t)|_{P{\to}\infty}$. As $\phi(r){\approx}{-}\gamma{+}{\ln}(r)$ for $r{\to}\infty$, one has
\begin{IEEEeqnarray}{rcl}
\epsilon(t)\stackrel{P{\to}\infty}{=}{\log_2}{\frac{1}{t}}&,\quad&
\kappa(t)\stackrel{P{\to}\infty}{=}-{\ln}(Pt){-}1{+}{\ln}4.\label{eq:phiapprox}
\end{IEEEeqnarray}
Substituting \eqref{eq:phiapprox} into \eqref{eq:dRt} yields
\begin{multline}
\Delta\tilde{R}_S^{eq}(t)|_{P{\to}\infty}{=}2{\log_2}\left(\frac{1}{t}{+}\frac{PM}{2(M{-}1)}2^{-\frac{B}{M{-}1}}\right){-}\\
{\log_2}\left(1{+}\frac{2}{t e}{-}\frac{2}{e}\right).\label{eq:dR_highP}
\end{multline}
By evaluating the first order derivative of \eqref{eq:dR_highP}, we can easily obtain
\begin{align}
t_S^{eq}{=}&\left\{\begin{array}{ll}\frac{1}{\frac{PM}{2(M{-}1)}2^{\frac{-B}{M{-}1}}{+}2{-}e} & \text{if }B{\leq}B_0^{eq};\\1&\text{if }B{>}B_0^{eq},\end{array}\right.\label{eq:t}
\end{align}
where $B_0^{eq}{=}(M{-}1)\left[{\log}_2\frac{PM}{2(M{-}1)}{-}{\log}_2(e{-}1)\right]$. Note that $B_0^{eq}$ acts as a threshold that switches the scheme from ZFBF with RVQ to RS-S if the feedback quality is not good enough.

Figure \ref{fig:RLM4B10} compares the analytical upper-bounds with the Monte Carlo simulation when $M{=}4$ and $B{=}10$. Specifically, for conventional ZFBF with quantized CSIT, the upper-bound $\Delta\tilde{R}_S^{eq}(1)$ is plotted by substituting $t{=}1$ into $\Delta\tilde{R}_S^{eq}(t)$, while the simulation is carried out with even power allocation. For RS-S, the upper-bound $\Delta\tilde{R}_S^{eq}(t_S^{eq})$ is plotted by substituting $t{=}t_S^{eq}$ into $\Delta\tilde{R}_S^{eq}(t)$. The simulations are carried out in two ways: 1) with an exhaustive (Ex) search for $t$; 2) with $t{=}t_S^{eq}$ in \eqref{eq:t}. We observe that $\Delta\tilde{R}_S^{eq}(t_S^{eq})$ is an upper-bound of the simulation result of the sum rate loss incurred by the RS-S scheme even though Assumption \ref{asp:sinr_approx} gives an upper-bound of ${\rm SINR}_c$. In addition, we can see that $t_S^{eq}$ in \eqref{eq:t} is a proper allocation for the RS-S scheme as the simulation of the RS-S scheme with $t_S^{eq}$ yields almost the same performance as the case with exhaustive search.

To gain insights into how the sum rate of RS-S scheme changes with $B$, let us substitute $t_S^{eq}$ into $\Delta\tilde{R}_S^{eq}(t)$ and evaluate $\Delta\tilde{R}_S^{eq}(t_S^{eq})$ focusing on high SNR and $B{<}B_0^{eq}$. It writes as
\begin{IEEEeqnarray}{rcl}
\!\!\!\!\Delta\tilde{R}_S^{eq}(t_S^{eq})|_{P{\to}\infty}{=}{\log}_2e{+}{\log}_2(\frac{PM}{(M{-}1)}2^{\frac{-B}{M-1}}{+}2{-}e).\label{eq:dR_highPt1}
\end{IEEEeqnarray}

\begin{myremark}\label{rmk:SNRgain1}
\textbf{(SNR gain offered by feedback quality increment:)}
We can see that in \eqref{eq:dR_highPt1}, a certain increment of $B$, equal to $b$ bits, results in $\frac{b}{M{-}1}$ bps/Hz sum rate enhancement for RS-S, where the term $2{-}e$ is negligible as we consider high SNR and $B$ is not a function of $P$. Such an enhancement can be interpreted as a $\frac{3b}{M{-}1}$ dB SNR gain for RS-S. Remarkably, this is an extraordinary distinction compared with ZFBF with RVQ and single-user transmission, i.e., TDMA.

For ZFBF with RVQ, the upper-bound of the sum rate loss writes as $\Delta\tilde{R}_S^{eq}(1){=}2{\log}_2(1{+}\frac{PM}{2(M{-}1)}2^{\frac{-B}{M{-}1}})$. Although increasing $B$ by $b$ bits yields a sum rate enhancement, it cannot be interpreted as a SNR gain because the sum rate saturates at high SNR. This can be seen from the pre-log factor of the $\Delta\tilde{R}_S^{eq}(1)$, which indicates a DoF loss of $2$. Similar observation was found in \cite{Jin06}.

For TDMA, an upper-bound on the sum rate, i.e., ${\log}_2(1{+}PM(1{-}2^{\frac{-B}{M{-}1}}))$, was shown in \cite{Jin06}. This indicates that increasing $B$ does not provide a significant gain especially when $B$ is already good enough.
\end{myremark}
By setting $M{=}4$ and different values of $B$, i.e., $10$ and $15$, Figure \ref{fig:RsM4B10} illustrates the simulation result of the ergodic sum rate of RS-S with $t_S^{eq}$ in \eqref{eq:t}. We see that, unlike ZFBF with RVQ, the sum rate of RS-S is increasing rather than saturating when $B$ does not change with $P$. The SNR gain stated in Remark \ref{rmk:SNRgain1} is verified as RS-S with $B{=}15$ yields a $5$dB SNR gain over the case with $B{=}10$ at high SNR regime. On the other hand, as mentioned in \cite{Jin06}, the saturation of the sum rate can be also avoided by doing TDMA. A thorough comparison between RS-S and TDMA will be presented in Section \ref{sec:performance}.

\subsection{A New Scaling Law of $B$}\label{sec:SJMBB}
\begin{figure}[t]
\renewcommand{\captionfont}{\small}
\captionstyle{center}
\centering
\subfigure[Overhead]{%{0.32\textwidth}
                \centering
                \includegraphics[width=0.22\textwidth,height=3.5cm]{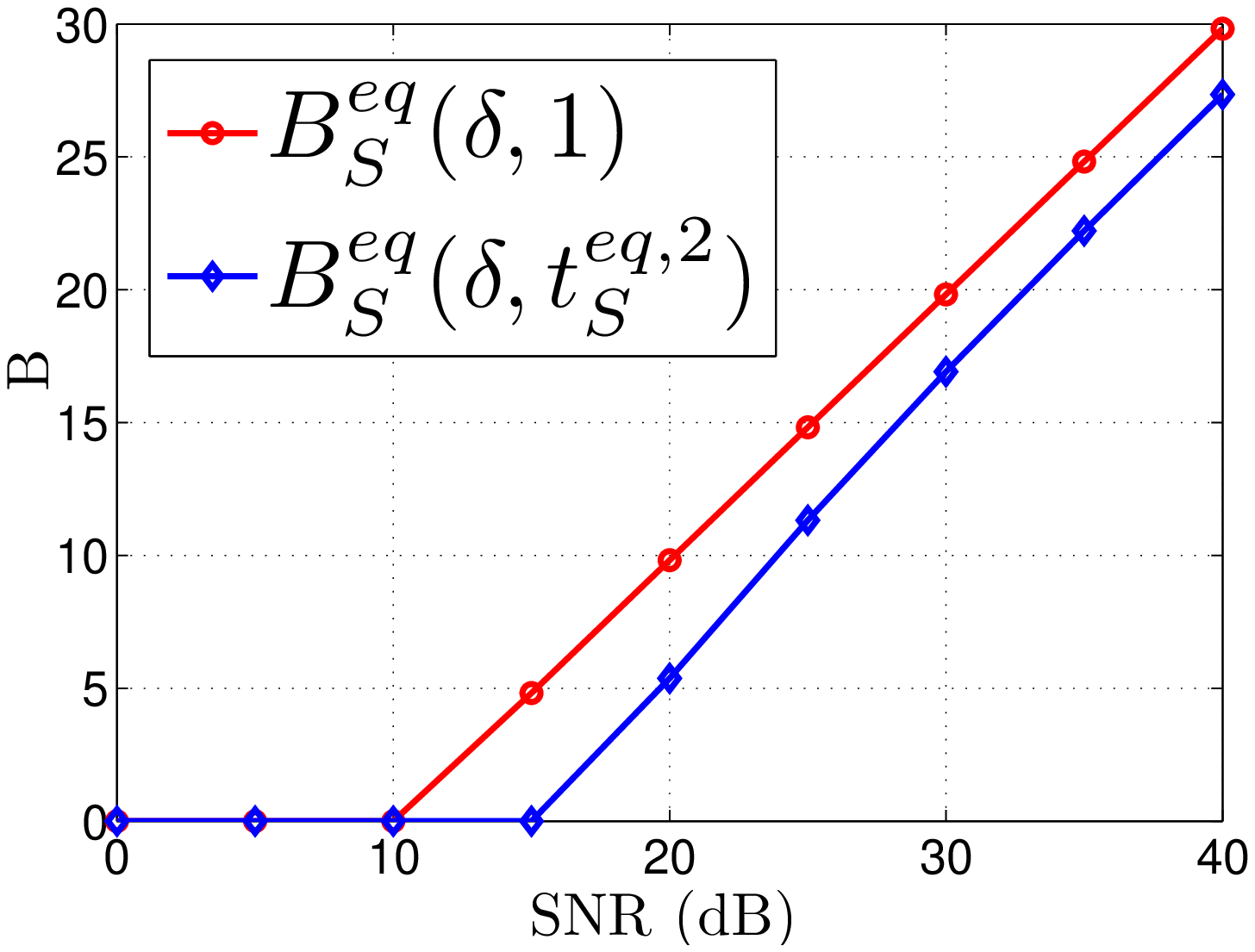}
                %\caption{$\mathcal{D}_1$}
                \label{fig:BM4b64}
        }%\end{subfigure}
\subfigure[Sum rate performance]{%{0.32\textwidth}
                \centering
                \includegraphics[width=0.26\textwidth,height=3.5cm]{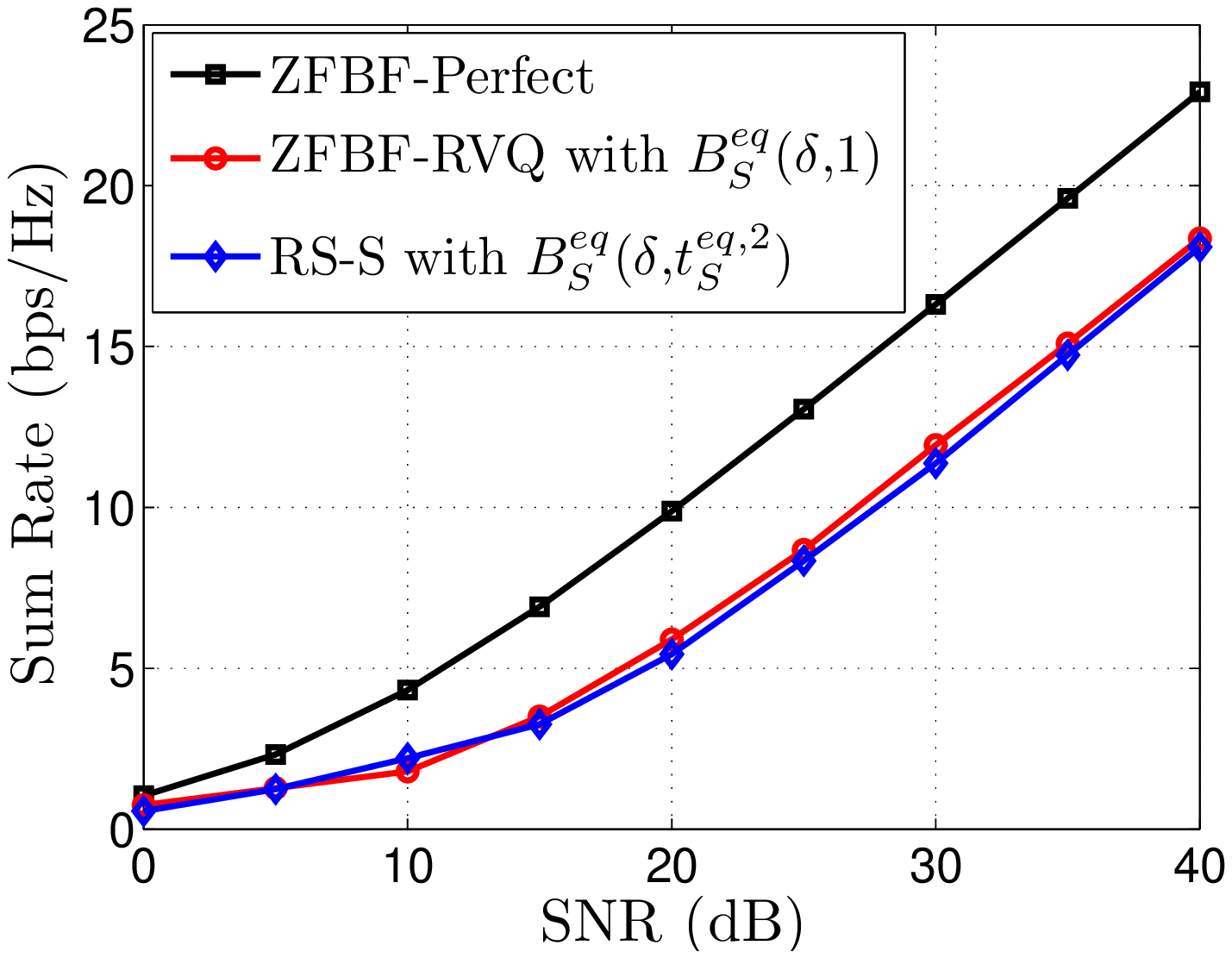}
                %\caption{$\mathcal{D}_1$}
                \label{fig:doubleBM4b32}
        }%\end{subfigure}
\caption{$M{=}4$, to achieve maximum ${\log}_2\delta{=}6$bps/Hz rate loss.}\label{fig:B}
\end{figure}
It has been shown in \cite{Jin06} that full DoF is achievable with ZFBF with RVQ if the number of feedback bits scales linearly with $M$ and SNR (in decibel). In this case, although RS-S scheme does not bring DoF gain over ZFBF with RVQ, it enables a feedback reduction to maintain a constant rate offset relative to the ZFBF with perfect CSIT. The following proposition specifies the scaling law of $B$ required by RS-S scheme to achieve a certain maximum allowable rate loss relative to the ZFBF with perfect CSIT.
\begin{myprop}\label{theo:B}
In the scenario where the two receivers have equal feedback qualities, to achieve a maximum allowable sum rate loss, equal to ${\log}_2\delta$ bps/Hz, relative to ZFBF with perfect CSIT, the number of feedback bits required by the RS-S scheme is given by $B_S^{eq}(\delta{,}t)$, where
\begin{multline}
B_S^{eq}(\delta{,}t){=}(M{-}1){\log}_2\frac{PM}{2(M{-}1)}
{-}\\(M{-}1){\log}_2\left(\frac{\sqrt{\delta}\sqrt{1{+}\frac{P(1{-}t)}{2}e^{\kappa(t)}}}
{t\cdot2^{\epsilon(t)}}{-}\frac{1}{t}\right),\label{eq:B1}
\end{multline}
where $\kappa(t)$, $\epsilon(t)$ and $t$ are the same as those introduced in Proposition \ref{theo:RateLossP1}.
\end{myprop}
\emph{Proof:} \eqref{eq:B1} is obtained as the inverse function of \eqref{eq:dRt}, namely setting $\Delta\tilde{R}_S^{eq}(t){=}{\log}_2\delta$ and calculating $B$ as a function of $\delta$ and $t$. $\hfill\Box$

We see that it is difficult to derive the optimal power splitting ratio $t^*{\triangleq}\arg\min_{0{<}t{\leq}1}B_S^{eq}(\delta{,}t)$ due to the complicated expression \eqref{eq:B1}. To gain insights into Proposition \ref{theo:B}, we obtain the optimal power splitting ratio at high SNR, namely $t_S^{eq{,}2}{\triangleq}\arg\min_{0{<}t{\leq}1}B_S^{eq}(\delta{,}t)|_{P{\to}\infty}$ where
\begin{multline}
B_S^{eq}(\delta{,}t)|_{P{\to}\infty}{=}(M{-}1){\log}_2P{-}(M{-}1){\log}_2\frac{2(M{-}1)}{M}
{-}\\(M{-}1){\log}_2\left(\sqrt{\delta\left(1{+}\frac{2}{te}{-}\frac{2}{e}\right)}{-}\frac{1}{t}\right).\label{eq:B_highP}
\end{multline}
By evaluating the first order derivative of \eqref{eq:B_highP}, we can easily obtain
\begin{align}
t_S^{eq{,}2}{=}&\left\{\begin{array}{ll}\frac{1}{\frac{\delta}{2e}{-}\frac{e}{2}{+}1} & \text{if }\delta{\geq}e^2,\\1&\text{if } 1{<}\delta{<}e^2.\end{array}\right.\label{eq:t2}
\end{align}
\begin{myremark}\label{rmk:dB1}
\textbf{[Feedback overhead reduction]} To achieve a maximum allowable rate loss equal to ${\log}_2\delta$ bps/Hz, by comparing the number of feedback bits required by RS-S, i.e., $B_S^{eq}(\delta{,}t_S^{eq{,}2})$, with that required by ZFBF with RVQ, i.e., $B_S^{eq}(\delta{,}1)$, one can compute the feedback overhead reduction as in \eqref{eq:deltaB1} at the top of next page.
\begin{figure*}
\begin{IEEEeqnarray}{rcl}
B_S^{eq}(\delta{,}1){-}B_S^{eq}(\delta{,}t_S^{eq{,}2})&{=}&
(M{-}1)\log_2\left[\left(\frac{\sqrt{\delta}\sqrt{1{+}\frac{P(1{-}t_S^{eq{,}2})}{2}e^{\kappa(t_S^{eq{,}2})}}}
{t_S^{eq{,}2}\cdot2^{\epsilon(t_S^{eq{,}2})}}{-}\frac{1}{t_S^{eq{,}2}}\right)/\left(\sqrt{\delta}{-}1\right)\right]\\
&{\stackrel{P{\to}\infty}{=}}&(M{-}1)\log_2\frac{\frac{\delta}{2e}{+}\frac{e}{2}{-}1}{\sqrt{\delta}{-}1}{,}\text{ for }
\delta{\geq}e^2.\label{eq:deltaB1}
\end{IEEEeqnarray}
\hrulefill
\end{figure*}
\end{myremark}

Setting the maximum allowable rate loss to be ${\log}_2\delta{=}6$ bps/Hz, we plot $B_S^{eq}(\delta{,}t_S^{eq{,}2})$ and $B_S^{eq}(\delta{,}1)$ in Figure \ref{fig:BM4b64} for $M{=}4$. Notably, at medium SNR ($15dB$), RS-S scheme requires $5$ bits less than ZFBF with RVQ. When it comes to high SNR, the feedback overhead reduction decreases to a constant. Figure \ref{fig:doubleBM4b32} illustrates the simulation result of the sum rate performance by applying $B_S^{eq}(\delta{,}1)$ to ZFBF with RVQ and $B_S^{eq}(\delta{,}t_S^{eq{,}2})$ to the RS-S scheme (the power splitting ratio in the simulation is $t_S^{eq{,}2}$), where ${\log}_2\delta{=}6$bps/Hz. Firstly, we see that both schemes achieve less than $6$bps/Hz rate loss relative to ZFBF with perfect CSIT with their respective scaling law of $B$. Secondly, both schemes achieve almost the same sum rate performance. This implies that Remark \ref{rmk:dB1} correctly characterizes the feedback overhead reduction offered by the RS-S scheme to achieve the same sum rate performance as ZFBF with RVQ.
%\footnote{Note that since $B(\delta{,}t)|_{P{\to}\infty}$ in \eqref{eq:B_highP} is the inverse function of $\Delta \tilde{R}(t)|_{P{\to}\infty}$ in \eqref{eq:dR_highP}, $t_S^{eq}{\triangleq}\arg\min_{0{<}t{\leq}1}\tilde{R}_S^{eq}(t)|_{P{\to}\infty}$ is equivalent with $t_S^{eq{,}2}{\triangleq}\arg\min_{0{<}t{\leq}1}B_S^{eq}(\delta{,}t)|_{P{\to}\infty}$} 
%\input{scn1_SJMB}

\section{RS-S and RS-ST with alternating receiver-specific feedback qualities}\label{sec:STJMB}
In this section, we focus on the scenario with alternating receiver-specific feedback qualities, where Rx1 utilizes $B_\beta$ (resp. $B_\alpha$) bits and Rx2 employs $B_\alpha$ (resp. $B_\beta$) to quantize their channels in channel use 1 (resp. 2). We firstly focus on the RS-S scheme and extend the results shown in the previous section. Secondly, we identify the benefit of the RS-ST scheme by comparing with the sum rate achieved with the RS-S scheme. For convenience, we use $\tau{\triangleq}B_{\beta}{-}B_{\alpha}$ (assuming $B_{\alpha}{<}B_\beta$) to represent the discrepancy between the feedback overhead employed by the two receivers in each channel use, and use $\bar{B}{\triangleq}\frac{B_{\alpha}{+}B_{\beta}}{2}$ to denote the average feedback overhead.
\subsection{Performing the RS-S scheme}\label{sec:SJMB2}
In this part, following the footsteps in the previous section, we study the sum rate performance of the RS-S scheme in the scenario with alternating receiver-specific feedback qualities. %Compared to the case where the two receivers have an equal feedback quality, its limitations are shown by the sum rate degradation and the increment of the average feedback overhead that is required to achieve a certain maximum allowable rate loss relative to ZFBF with perfect CSIT.
\subsubsection{Sum rate loss}
Let us use $\Delta R_S^{rs}(t)$ to denote the sum-rate loss incurred by the RS-S scheme in the scenario with alternating receiver-specific feedback qualities. As the sum rate achieved by RS-S scheme in channel use 1 and 2 are statistically equivalent, we only focus on the sum rate achieved in a single channel use. Reusing the proof of Proposition \ref{theo:RateLossP1}, an upper-bound of $\Delta R_S^{rs}(t)$ is given below.
\begin{myprop}{\label{prop:ratelossQ1}}
In the scenario with alternating receiver-specific feedback qualities, the sum-rate loss incurred by the RS-S scheme with RVQ relative to the ZFBF with perfect CSIT is upper-bounded by
\begin{multline}
\Delta R_S^{rs}(t){\leq}\Delta\tilde{R}_S^{rs}(t){=}2\epsilon(t){+}{\log}_2(1{+}t\Lambda_{\alpha}){+}{\log}_2(1{+}t\Lambda_{\beta})
{-}\\{\log}_2(1{+}\frac{P(1{-}t)}{2}e^{\kappa(t)}),\label{eq:RLQ1}
\end{multline}
where $\Lambda_{\alpha}{=}\frac{PM}{2(M{-}1)}2^{\frac{-B_{\alpha}}{M-1}}$, $\Lambda_{\beta}{=}\frac{PM}{2(M{-}1)}2^{\frac{-B_{\beta}}{M-1}}$, while $\kappa(t)$, $\epsilon(t)$ and $t$ are the same as those introduced in Proposition \ref{theo:RateLossP1}.
\end{myprop}

Similar to the analysis in the previous section, it is difficult to obtain a closed-form solution of $t^*{\triangleq}\arg\min_{0{<}t{\leq}1}\Delta\tilde{R}_S^{rs}(t)$ for arbitrary SNR due to the complicated expression of $\Delta\tilde{R}_S^{rs}(t)$. Hence, we calculate an optimal power splitting ratio that minimizes $\Delta\tilde{R}_S^{rs}(t)$ at high SNR, namely $t_S^{rs}{\triangleq}\arg\min_{0{<}t{\leq}1}\Delta\tilde{R}_S^{rs}(t)|_{P{\to}\infty}$. Specifically, $\Delta\tilde{R}_S^{rs}(t)|_{P{\to}\infty}$ writes as
\begin{multline}
\Delta\tilde{R}_S^{rs}(t)|_{P{\to}\infty}{=}
\log_2\left(\frac{1}{t}{+}\Lambda_{\alpha}\right){+}\log_2\left(\frac{1}{t}{+}\Lambda_{\beta}\right){-}\\
\log_2\left(1{+}\frac{2}{te}{-}\frac{2}{e}\right).\label{eq:dRQ1_highP}
\end{multline}
Then, by evaluating the first order derivative of \eqref{eq:dRQ1_highP}, it can be shown that
\begin{equation}
t_S^{rs}{=}\left\{\begin{array}{ll}1&\bar{B}{\geq}\bar{B}_0^{rs}(\Theta);\\
\frac{1}{\sqrt{(\Lambda_{\alpha}{-}\frac{e{-}2}{2})(\Lambda_{\beta}{-}\frac{e{-}2}{2})}{-}\frac{e{-}2}{2}}&
\bar{B}{<}\bar{B}_0^{rs}(\Theta),\end{array}\right.\label{eq:t3}
\end{equation}
where $\bar{B}_0^{rs}(\Theta)$ is given in \eqref{eq:B02} at the top of next page, and
\begin{figure*}
\begin{IEEEeqnarray}{rcl}
\bar{B}_0^{rs}(\Theta)&{=}&(M{-}1){\log}_2P{-}(M{-}1){\log}_2\frac{2(M{-}1)}{M}{-}
(M{-}1){\log}_2\left(\sqrt{\frac{e^2}{4}{+}(e{-}2)^2\frac{\Theta(\Theta{-}4)}{16}}{+}\frac{e{-}2}{4}(\Theta{-}2)\right),
\label{eq:B02}
\end{IEEEeqnarray}
\hrulefill
\end{figure*}
\begin{IEEEeqnarray}{rcl}
\Theta&{=}&2^{\frac{-\tau}{2(M{-}1)}}{+}2^{\frac{\tau}{2(M{-}1)}}{+}2.\label{eq:theta}
\end{IEEEeqnarray}
Note that $\bar{B}_0^{rs}(\Theta)$ is the threshold that switches the scheme between RS-S and ZFBF with RVQ. Clearly, $\bar{B}_0^{rs}(\Theta)$ is monotonically decreasing with $\Theta$, i.e., $\tau$. When $B_{\alpha}{=}B_{\beta}$, we have $\bar{B}_0^{rs}(\Theta){=}B_0^{eq}$ and \eqref{eq:RLQ1} and \eqref{eq:t3} become \eqref{eq:dRt} and \eqref{eq:t}, respectively.

To gain insights into the impact of having receiver-specific feedback qualities, let us consider $\bar{B}{<}\bar{B}_0^{rs}(\Theta)$ and plug $t_S^{rs}$ into \eqref{eq:dRQ1_highP}. The upper-bound of the sum rate loss at high SNR can be derived as
\begin{IEEEeqnarray}{rrl}
&\Delta\tilde{R}_S^{rs}(t_S^{rs})&|_{P{\to}\infty}{=}\nonumber\\
&&{\log}_2\left(t_S^{rs}(\sqrt{\Lambda_{\alpha}\Lambda_{\beta}}{-}\frac{1}{t_S^{rs}})^2{+}
(\sqrt{\Lambda_{\alpha}}{+}\sqrt{\Lambda_{\beta}})^2\right){-}\nonumber\\
&&\log_2\left(\frac{2}{e}{+}(1{-}\frac{2}{e})t_S^{rs}\right).\label{eq:dRQ1_highP1}
\end{IEEEeqnarray}
Note that it is cumbersome to quantify the term $t_S^{rs}(\sqrt{\Lambda_{\alpha}\Lambda_{\beta}}{-}\frac{1}{t_S^{rs}})^2$ due to the constant terms $\frac{e{-}2}{2}$ in $t_S^{rs}$. Thus, to obtain a quantitative result, we further upper-bound $\Delta\tilde{R}_S^{rs}(t_S^{rs})|_{P{\to}\infty}$ by $\Delta\tilde{R}_S^{rs}(\tilde{t}_S^{rs})|_{P{\to}\infty}$, where $\tilde{t}_S^{rs}{\triangleq}\frac{1}{\sqrt{\Lambda_\alpha\Lambda_\beta}}$, because $t_S^{rs}{\triangleq}\arg\min_{0{<}t{\leq}1}\Delta\tilde{R}_S^{rs}(t)|_{P{\to}\infty}$. Specifically,
\begin{IEEEeqnarray}{rrl}
\!\!\!\!\!\!\!&\Delta\tilde{R}_S^{rs}(t_S^{rs})&|_{P{\to}\infty}{\leq}
\Delta\tilde{R}_S^{rs}(\tilde{t}_S^{rs})|_{P{\to}\infty}\nonumber\\
\!\!\!\!\!\!\!&{=}&{\log}_2(\sqrt{\Lambda_{\alpha}}{+}\sqrt{\Lambda_{\beta}})^2{-}
\log_2\left(\frac{2}{e}{+}(1{-}\frac{2}{e})\tilde{t}_S^{rs}\right)\label{eq:dRQ1_highP2}\\
\!\!\!\!\!\!\!&{\leq}&{\log}_2\left(\frac{PM\cdot2^{\frac{\bar{B}}{M{-}1}}}{2(M{-}1)}\cdot\Theta\right){+}\log_2\frac{e}{2}.\label{eq:dRQ1_highP3}
\end{IEEEeqnarray}
Note that by comparing \eqref{eq:dRQ1_highP3} with \eqref{eq:dRQ1_highP1}, we can see that \eqref{eq:dRQ1_highP3} upper-bounds $\Delta\tilde{R}_S^{rs}(t_S^{rs})|_{P{\to}\infty}$ within ${\log}_2\frac{e}{2}{\approx}0.44$. This is because the first term in \eqref{eq:dRQ1_highP1} is greater than the first term in \eqref{eq:dRQ1_highP3} while the second term ${\log}_2\frac{e}{2}{\approx}0.44$ in \eqref{eq:dRQ1_highP3} upper-bounds the second term in \eqref{eq:dRQ1_highP1}.
\begin{myremark}\label{rmk:SJMBdegrade}
\textbf{(Sum rate degradation of RS-S scheme with alternating receiver-specific feedback qualities:)} From \eqref{eq:dRQ1_highP3}, we can see that, compared to the case $\tau{=}0$ where $\Theta{=}4$, the sum rate degradation incurred by $\tau{>}0$ can be characterized by ${\log}_2\frac{\Theta}{4}$ bps/Hz. Similar to Remark \ref{rmk:SNRgain1}, this degradation can be interpreted as a $3{\log}_2\frac{\Theta}{4}{\approx}3(\frac{\tau}{2(M{-}1)}{-}2)$ dB SNR loss, if $\tau$ is relatively large.
\end{myremark}

\subsubsection{Scaling law of $\bar{B}$}
Inverting \eqref{eq:RLQ1} with the respect of (w.r.t.) $\bar{B}$ yields the following proposition.
\begin{myprop}\label{coro:BSrs}
In the scenario with alternating receiver-specific feedback qualities, to achieve a maximum allowable rate loss, equal to ${\log}_2\delta$ bps/Hz, relative to ZFBF with perfect CSIT, the average number of feedback bits required by the RS-S scheme is given by
\begin{IEEEeqnarray}{rrl}
\!\!\!\!\!\!&\bar{B}_S^{rs}(\delta{,}t){=}&(M{-}1){\log}_2\frac{PM}{2(M{-}1)}{-}\nonumber\\
\!\!\!\!\!\!&(M{-}1){\log}_2&\left(\sqrt{\frac{\Theta^2{-}4\Theta}{4t^2}{+}\frac{\delta\left(
1{+}\frac{P(1{-}t)}{2}e^{\kappa(t)}\right)}
{t^2\cdot2^{2\epsilon(t)}}}{-}\frac{\Theta{-}2}{2t}\right)\!\!,\label{eq:B2}
\end{IEEEeqnarray}
where $\Theta$ is given by \eqref{eq:theta}, and $\kappa(t)$, $\epsilon(t)$ and $t$ are the same as those introduced in Proposition \ref{theo:RateLossP1}.
\end{myprop}
\emph{Proof:} see Appendix D. $\hfill\Box$

Similar to the previous analysis, it is difficult to calculate the optimal power splitting ratio $t^*{\triangleq}\arg\min_{0{<}t{\leq}1} \bar{B}_S^{rs}(\delta{,}t)$ for arbitrary SNR. However, we can make progress at high SNR regime. In particular, when $P{\to}\infty$, the optimal power splitting ratio is defined as $t_S^{rs{,}2}{\triangleq}\min_{0{<}t{\leq}1}\bar{B}_S^{rs}(\delta{,}t)|_{P{\to}\infty}$. With the derivation presented in Appendix D, we obtain $t_S^{rs{,}2}$ in \eqref{eq:t4} at the top of next page.
\begin{figure*}
\begin{IEEEeqnarray}{rcl}
t_S^{rs{,}2}&{=}&\left\{\begin{array}{ll}\left[\sqrt{\frac{4(\Theta{-}2)^2}{e^2(\Theta^2{-}4\Theta)^2}\delta^2
{-}\frac{(\Theta{-}2)^2}{\Theta^2{-}4\Theta}\delta(1{-}\frac{2}{e})}{-}\frac{4\delta}{e(\Theta^2{-}4\Theta)}\right]^{-1}
&\delta{>}\delta_0(\Theta){;}\\1&1{<}\delta{\leq}\delta_0(\Theta){;}
\end{array}\right.\label{eq:t4}\\
\delta_0(\Theta)&{\triangleq}&\sqrt{\frac{e^2}{4}(\Theta^2{-}4\Theta){+}\left(\frac{e^2}{8}(\Theta{-}2)^2(1{-}\frac{2}{e}){+}e\right)^2}
{+}\frac{e^2}{8}(\Theta{-}2)^2(1{-}\frac{2}{e}){+}e
\end{IEEEeqnarray}
\hrulefill
\end{figure*}
Note that in Appendix D, $t_S^{rs{,}2}$ is the solution to a quadratic formula. When $B_\alpha{=}B_\beta$, the quadratic formula degrades to a linear formula, where the resultant $t_S^{rs{,}2}$ becomes $t_S^{eq{,}2}$ in \eqref{eq:t2} and the threshold $\delta_0(\Theta)$ becomes $e^2$. Clearly, $\delta_0(\Theta)$ is monotonically increasing with $\Theta$ (or $\tau$) as $\Theta{\geq}4$. This indicates that the threshold where S-JMB starts to offer feedback reduction over ZFBF with RVQ grows with $\Theta$ (or $\tau$).
\begin{figure}[t]
\renewcommand{\captionfont}{\small}
\captionstyle{center}
\centering
\includegraphics[width=0.3\textwidth,height=4cm]{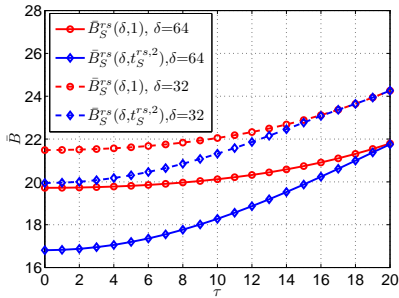}
\caption{$\bar{B}$ vs. $\tau$, $M{=}4$ and $P{=}30$dB.}\label{fig:Bvstau}
\end{figure}
\begin{myremark}\label{rmk:BSJMB2}
\textbf{(Average feedback overhead reduction offered by the RS-S scheme over ZFBF with RVQ when there are alternating receive-specific feedback qualities:)} To achieve a maximum allowable sum rate loss relative to ZFBF with perfect CSIT, equal to ${\log}_2\delta$ bps/Hz, compared to the feedback overhead required by ZFBF with RVQ (i.e., $B_S^{rs}(\delta{,}1)$), performing RS-S scheme enables an average feedback overhead reduction that scales as
\begin{multline}
\bar{B}_S^{rs}(\delta{,}1){-}\bar{B}_S^{rs}(\delta{,}t_S^{rs{,}2}){\stackrel{P{\to}\infty}{=}}\\(M{-}1){\log}_2
\frac{\sqrt{\frac{\Theta^2{-}4\Theta}{4(t_S^{rs{,}2})^2}{+}\delta\left(1{+}\frac{2}{t_S^{rs{,}2}e}{-}\frac{2}{e}\right)}{-}
\frac{\Theta{-}2}{2t_S^{rs{,}2}}}{\sqrt{\frac{\Theta^2{-}4\Theta}{4}{+}\delta}{-}\frac{\Theta{-}2}{2}}.\label{eq:dB2}
\end{multline}
\end{myremark}
Due to the complicated expression of \eqref{eq:t4} and \eqref{eq:dB2}, it is cumbersome to analyze the impact of $\tau$ on the feedback overhead reduction. In Figure \ref{fig:Bvstau}, we plot $\bar{B}_S^{rs}(\delta{,}1)$ and $\bar{B}_S^{rs}(\delta{,}t_S^{rs{,}2})$ for $M{=}4$ and $P{=}30$dB. When $\tau$ increases, we can see that the gap between $\bar{B}_S^{rs}(\delta{,}1)$ and $\bar{B}_S^{rs}(\delta{,}t_S^{rs{,}2})$ decreases. This indicates that the feedback overhead reduction in \eqref{eq:dB2} offered by the RS-S scheme over ZFBF with RVQ is decreasing with $\tau$. When $\delta_0(\Theta)$ is equal to $\delta$, $\bar{B}_S^{rs}(\delta{,}1)$ and $\bar{B}_S^{rs}(\delta{,}t_S^{rs{,}2})$ coincide. %Specifically, when $\tau{=}17$, one has $\delta_0(\Theta){\approx}34$ and $\bar{B}_S^{rs}(32{,}1)$ and $\bar{B}_S^{rs}(32{,}t_S^{rs{,}2})$ coincide; when $\tau{=}20$, one has $\delta_0(\Theta){\approx}60$ and $\bar{B}_S^{rs}(64{,}1)$ and $\bar{B}_S^{rs}(64{,}t_S^{rs{,}2})$ nearly coincide.

\subsection{Benefit of the Space-Time Transmission}
\begin{figure}[t]
\renewcommand{\captionfont}{\small}
\captionstyle{center}
\centering
\subfigure[Rate loss, $\tau{=}6$.]{%{0.32\textwidth}
                \centering
                \includegraphics[width=0.3\textwidth,height=4cm]{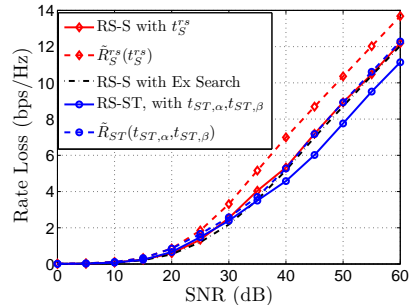}
                %\caption{$\mathcal{D}_1$}
                \label{fig:RLSTJMB}
        }\\%\end{subfigure}
\subfigure[Sum rate, $\tau{=}6$ and $\tau{=}10$.]{%{0.32\textwidth}
                \centering
                \includegraphics[width=0.3\textwidth,height=4cm]{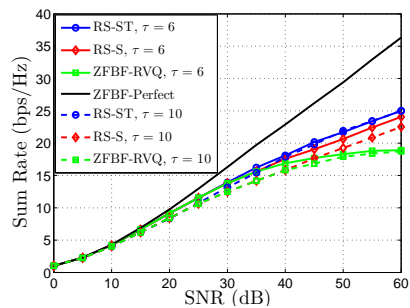}
                %\caption{$\mathcal{D}_1$}
                \label{fig:STvsS1}
        }%\end{subfigure}
\caption{Simulation results for RS-ST, RS-S and ZFBF with $\bar{B}{=}10$ and $M{=}2$.}
\end{figure}
Previous subsection identifies the impact of $\tau$ on the sum rate performance and the feedback overhead reduction over ZFBF with RVQ achieved by the RS-S scheme in the scenario with alternating receiver-specific CSIT qualities. In this part, we analyze the sum rate performance of the RS-ST scheme. By comparing with the RS-S scheme, we will show the benefit of transmitting an additional common message, i.e., $c_0$, using a space-time transmission.
\subsubsection{Sum rate loss}
Recalling the SINR expressions in \eqref{eq:SINRst}, we see that the SINR of $c_1$, $u_{11}$ and $u_{21}$ are statistically equivalent with that of $c_2$, $u_{22}$ and $u_{12}$, respectively. Hence, the sum rate achieved by RS-ST writes as
\begin{multline}
R_{ST}(t_\alpha{,}t_\beta){\triangleq} \frac{1}{2}(2R_{u{11}}(t_\alpha{,}t_\beta){+}2R_{u{21}}(t_\alpha{,}t_\beta){+}\\2R_{c1}(t_\alpha{,}t_\beta){+}R_{c0}(t_\alpha{,}t_\beta))\nonumber
\end{multline}
Moreover, the sum rate achieved by ZFBF with perfect CSIT are statistically equivalent in these two channel uses. Thus, let us write the sum rate achieved by ZFBF with perfect CSIT as $R_{11}^p{+}R_{21}^p$, where $R_{11}^p$ and $R_{21}^p$ denote the rate achieved by the private messages intended for Rx1 and Rx2 in channel use 1, respectively. Consequently, the sum rate loss incurred by the RS-ST scheme relative to ZFBF with perfect CSIT is defined as $\Delta R_{ST}^{rs}(t_{\alpha}{,}t_{\beta}){\triangleq}R_{11}^p{+}R_{21}^p{-}R_{ST}(t_\alpha{,}t_\beta)$. An upper-bound of $\Delta R_{ST}^{rs}(t_{\alpha}{,}t_{\beta})$ is stated below.

\begin{myprop}\label{prop:RateLossP2}
In the scenario with alternating receiver-specific feedback qualities, the sum-rate loss incurred by the RS-ST scheme with RVQ relative to the ZFBF with perfect CSIT is upper-bounded as
\begin{IEEEeqnarray}{rcl}
\!\!\!\!\!\!\Delta R_{ST}(t_{\alpha}{,}t_{\beta})&{\leq}&\Delta\tilde{R}_{ST}(t_{\alpha}{,}t_{\beta})\nonumber\\
\!\!\!\!\!\!&{=}&\mu(t_\alpha{,}t_\beta){-}\varrho(t_\alpha{,}t_\beta){+}\!\log_2\!\left(1{+}t_\alpha\Lambda_{\alpha}\right){+}\nonumber\\
\!\!\!\!\!\!&&\log_2\!\left(1{+}t_\beta\Lambda_{\beta}\right)\!{-}\!\log_2\!\left[\!1{+}\frac{P(1{-}t_\beta)}{2}e^{\kappa(t_\beta)}\!\right]\!\!,
\label{eq:dRP2}
\end{IEEEeqnarray}
with $\mu(t_\alpha{,}t_\beta){=}
\frac{1}{\ln2}\left[\phi(\frac{P}{2}){-}\phi(\frac{Pt_\beta}{2}){+}\phi(\frac{P}{2}){-}\phi(\frac{Pt_\alpha}{2})\right]$ and $\varrho(t_\alpha{,}t_\beta){=}\frac{1}{\ln2}\left[\phi(\frac{Pt_\beta}{2}){-}\frac{\phi(\frac{Pt_\beta}{4})}{2}
{-}\phi(\frac{Pt_\alpha}{2}){+}\frac{\phi(\frac{Pt_\alpha}{4})}{2}\right]$, while $\Lambda_{\alpha}$, $\Lambda_{\beta}$, $\kappa(t_\beta)$, $t_\alpha{\in}(0{,}t_\beta]$ and $t_\beta{\in}[t_\alpha{,}1]$ are the same as those introduced in Proposition \ref{prop:ratelossQ1}.
\end{myprop}
\emph{Proof:} see Appendix E. $\hfill\Box$

In \eqref{eq:dRP2}, $\mu(t_\alpha{,}t_\beta)$ refers to the rate loss incurred by the power decrement of the private messages, while $\varrho(t_\alpha{,}t_\beta)$ characterizes the rate achieved by $c_0$, which is essentially determined by the discrepancy between $t_{\alpha}$ and $t_{\beta}$. The rate loss incurred by the ZF precoders in RS-ST with RVQ is shown by $\log_2\left(1{+}t_\beta\Lambda_{\beta}\right)$ and $\log_2\left(1{+}t_\alpha\Lambda_{\alpha}\right)$. The last term represents the rate achieved by $c_1$ and $c_2$.

Note that $\Delta\tilde{R}_{ST}(t_{\alpha}{,}t_{\beta})$ in \eqref{eq:dRP2} gives an upper-bound for arbitrary SNR, but it is difficult to get explicit insights from $\Delta\tilde{R}_{ST}(t_{\alpha}{,}t_{\beta})$. Moreover, after writing  $\Delta\tilde{R}_{ST}(t_{\alpha}{,}t_{\beta})$ at high SNR, namely
\begin{multline}
\Delta\tilde{R}_{ST}(t_{\alpha}{,}t_{\beta})|_{P{\to}\infty}{=}
{\log}_2\left(\frac{1}{t_{\beta}}{+}\Lambda_{\beta}\right){+}{\log}_2\left(\frac{1}{t_{\alpha}}{+}\Lambda_{\alpha}\right)
{-}\\ \frac{1}{2}{\log}_2\frac{t_{\beta}}{t_{\alpha}}{-}{\log}_2\left(1{+}\frac{2}{t_{\beta}e}{-}\frac{2}{e}\right),
\label{eq:dRP2_highP}
\end{multline}
we find that it is still cumbersome to calculate the closed-form solution of the power splitting ratios $(t_{\alpha}^*{,}t_{\beta}^*){\triangleq}\arg\min_{0{<}t_\alpha{\leq}t_\beta{\leq}1}\Delta\tilde{R}_{ST}(t_\alpha{,}t_\beta)|_{P{\to}\infty}$ as \eqref{eq:dRP2_highP} is a non-convex function of two arguments. Hence, for the sake of analysis tractability, we choose $t_\alpha$ and $t_\beta$ as
\begin{equation}
t_{ST{,}\beta}{=}\min\{\Lambda_{\beta}^{-1}{,}1\},\quad t_{ST{,}\alpha}{=}\min\{\Lambda_{\alpha}^{-1}{,}1\}.\label{eq:tab}
\end{equation}
Note that \eqref{eq:tab} follows the power allocation in \cite{Elia13}, where the power allocated to the private messages is chosen to ensure that the residual interference after ZFBF with imperfect CSIT is drowned by the noise, i.e., $\Lambda_{\alpha}t_{\alpha}{\leq}1$ and $\Lambda_{\beta}t_{\beta}{\leq}1$. Although $t_{ST{,}\alpha}$ and $t_{ST{,}\beta}$ in \eqref{eq:tab} are non-optimal in minimizing the rate loss, they provide a baseline of how much benefit we can gain from the space-time transmission of $c_0$.

In Figure \ref{fig:RLSTJMB}, we firstly compare the simulation results of RS-S with $t{=}t_S^{rs}$ in \eqref{eq:t3} and RS-ST with $t_\alpha{=}t_{ST{,}\alpha}$ and $t_\beta{=}t_{ST{,}\beta}$ in \eqref{eq:tab}, with their corresponding analytical upper-bounds, namely $\Delta\tilde{R}_S^{rs}(t_S^{rs})$ and $\Delta\tilde{R}_{ST}(t_{ST{,}\alpha}{,}t_{ST{,}\beta})$, respectively. We can see that Proposition \ref{prop:ratelossQ1} and \ref{prop:RateLossP2} upper-bound the sum rate loss incurred by the RS-S and RS-ST scheme in the scenario with alternating receiver-specific feedback qualities, even though Assumption \ref{asp:sinr_approx} provides an upper-bound of the SINR of the common messages. Secondly, we also plot the simulation results of the RS-S scheme with an optimal power splitting ratio obtained by exhaustive search. We can see that the simulation of the RS-S scheme with $t_S^{rs}$ in \eqref{eq:t3} yields almost the same performance as the case with exhaustive search. This indicates that $t_S^{rs}$ is a proper allocation for the RS-S scheme in the scenario with alternating receiver-specific feedback qualities.

Next, to obtain insights into the sum rate performance of the RS-ST scheme, we plug $t_\beta{=}t_{ST{,}\beta}$ and $t_\alpha{=}t_{ST{,}\alpha}$ into \eqref{eq:dRP2_highP} and derive $\Delta\tilde{R}_{ST}(t_{ST{,}\alpha}{,}t_{ST{,}\beta})|_{P{\to}\infty}$ as
\begin{IEEEeqnarray}{rrl}
&\Delta\tilde{R}_{ST}&(t_{ST{,}\alpha}{,}t_{ST{,}\beta})|_{P{\to}\infty}\nonumber\\
&{=}&{\log}_2\left(\frac{1}{t_{ST{,}\beta}}{+}\Lambda_{\beta}\right){+}{\log}_2\left(\frac{1}{t_{ST{,}\alpha}}{+}\Lambda_{\alpha}\right)
{+}\nonumber\\&&\frac{1}{2}{\log}_2t_{ST{,}\beta}t_{ST{,}\alpha}{-}{\log}_2(\frac{2}{e}{+}(1{-}\frac{2}{e})t_{ST{,}\beta}),\nonumber\\
&{\leq}&{\log}_2\frac{PM\cdot2^{\frac{-\bar{B}}{M{-}1}}}{2(M{-}1)}{+}2{+}{\log}_2\frac{e}{2},\label{eq:dRP2_highP1}
\end{IEEEeqnarray}
where the inequality is because $\frac{2}{e}{+}(1{-}\frac{2}{e})t_{ST{,}\beta}{\geq}\frac{2}{e}$. In \eqref{eq:dRP2_highP1}, we see that the sum rate loss incurred by the the RS-ST scheme at high SNR is not a function of $\Theta$. This indicates that, when $\bar{B}$ is fixed, performing RS-ST scheme produces similar results for any choice of $B_{\alpha}$ and $B_{\beta}$. This is in contrast to the RS-S scheme, where the sum rate degrades dramatically with $\tau$.
\begin{myremark}\label{rmk:SNRgainST}
\textbf{[SNR gain offered by RS-ST over RS-S scheme:]}
Using \eqref{eq:dRQ1_highP3} and \eqref{eq:dRP2_highP1}, we quantify the gap between the sum rate achieved by RS-ST and RS-S scheme at high SNR as
\begin{multline}
\Delta\tilde{R}_{S}(t_S^{rs})|_{P{\to}\infty}{-}\Delta\tilde{R}_{ST}(t_{ST{,}\alpha}{,}t_{ST{,}\beta})|_{P{\to}\infty}
{\approx}\\2{\log}_2\frac{2^{\frac{-\tau}{4(M{-}1)}}{+}2^{\frac{\tau}{4(M{-}1)}}}{2},
\end{multline}
where the approximation is due to the fact that \eqref{eq:dRQ1_highP3} and \eqref{eq:dRP2_highP1} are upper-bounds of $\Delta\tilde{R}_{S}(t_S^{rs})|_{P{\to}\infty}$ and $\Delta\tilde{R}_{ST}(t_{ST{,}\alpha}{,}t_{ST{,}\beta})|_{P{\to}\infty}$, respectively. For a large value of $\tau$, we can see that RS-ST scheme offers a SNR gain of $3(\frac{\tau}{2(M{-}1)}{-}2)$ dB over the RS-S scheme, which indicates that in the scenario with alternating receiver-specific feedback qualities, the sum rate degradation of the RS-S scheme mentioned in Remark \ref{rmk:SJMBdegrade} can be avoided by the space-time transmission of the additional common message, i.e., $c_0$.
\end{myremark}

Figure \ref{fig:STvsS1} illustrates the simulation results of ZFBF with RVQ, RS-S with $t_S^{rs}$ in \eqref{eq:t3} and RS-ST scheme with $t_{ST{,}\alpha}$ and $t_{ST{,}\beta}$ in \eqref{eq:tab} for $M{=}2$, $\bar{B}{=}20$ and different values of $\tau$. We can see that RS-ST scheme with $\tau{=}6$ and $\tau{=}10$ yield the same performance at high SNR, offering $2{\sim}3$dB and $8{\sim}9$dB SNR gain over RS-S scheme when $\tau{=}6$ and $\tau{=}10$ respectively.
\subsubsection{Scaling law of $\bar{B}$}
Let $\bar{B}_{ST}(\delta)$ denote the average feedback overhead required by the RS-ST scheme to achieve a maximum allowable rate loss, equal to ${\log}_2\delta$ bps/Hz, relative to ZFBF with perfect CSIT. To characterize $\bar{B}_{ST}(\delta)$, we can invert \eqref{eq:dRP2} w.r.t. $\bar{B}$ and evaluate the resulted inverse function at high SNR. But these footsteps will result in an equation similar to \eqref{eq:dRP2_highP}, which does not allow us to find a closed-form solution of the optimal power splitting ratios. Instead, using \eqref{eq:dRP2_highP1} and \eqref{eq:dRQ1_highP3}, we will firstly obtain a tractable result of the average feedback overhead reduction enabled by RS-ST over RS-S, such that both RS-ST and RS-S schemes achieve the same sum rate performance. Secondly, using such a quantity, we find a tractable expression of $\bar{B}_{ST}(\delta)$ as a function of the average feedback overhead required by the RS-S scheme to maintain ${\log}_2\delta$ bps/Hz sum rate offset relative to ZFBF with perfect CSIT.

Specifically, when $\tau$ is fixed, by setting \eqref{eq:dRP2_highP1} equal to \eqref{eq:dRQ1_highP3} and through some simple manipulations, we can calculate that the difference between the average number of feedback bits employed by RS-S in \eqref{eq:dRQ1_highP3} and that employed by RS-ST in \eqref{eq:dRP2_highP1} scales as
\begin{equation}
(M{-}1){\log}_2\frac{\Theta}{4}{=}2(M{-}1){\log}_2\frac{2^{\frac{-\tau}{4(M{-}1)}}{+}2^{\frac{\tau}{4(M{-}1)}}}{2}.\label{eq:BSTminusBS}
\end{equation}
If $\tau$ is relatively large (w.r.t $4(M{-}1)$), such a feedback overhead reduction offered by RS-ST over RS-S writes as $\frac{\tau}{2}{-}2(M{-}1)$. This quantity allows us to express $\bar{B}_{ST}(\delta)$ as
\begin{IEEEeqnarray}{rcl}
\!\!\!\!\!\!\bar{B}_{ST}(\delta)&{=}&\bar{B}_S^{rs}(\delta{,}t_S^{rs{,}2}){-}
2(M{-}1){\log}_2\frac{2^{\frac{-\tau}{4(M{-}1)}}{+}2^{\frac{\tau}{4(M{-}1)}}}{2},\label{eq:BST}
\end{IEEEeqnarray}
where $\bar{B}_S^{rs}(\delta{,}t_S^{rs{,}2})$ in \eqref{eq:B2} with $t_S^{rs{,}2}$ in \eqref{eq:t4} characterizes the average number of feedback bits required by the RS-S scheme to achieve maximum ${\log}_2\delta$ bps/Hz rate loss relative to ZFBF with perfect CSIT in the scenario with alternating receiver-specific feedback qualities.

\begin{figure}[t]
\renewcommand{\captionfont}{\small}
\captionstyle{center}
\centering \subfigure[Average Overhead]{%{0.32\textwidth}
                \centering
                \includegraphics[width=0.22\textwidth,height=3.5cm]{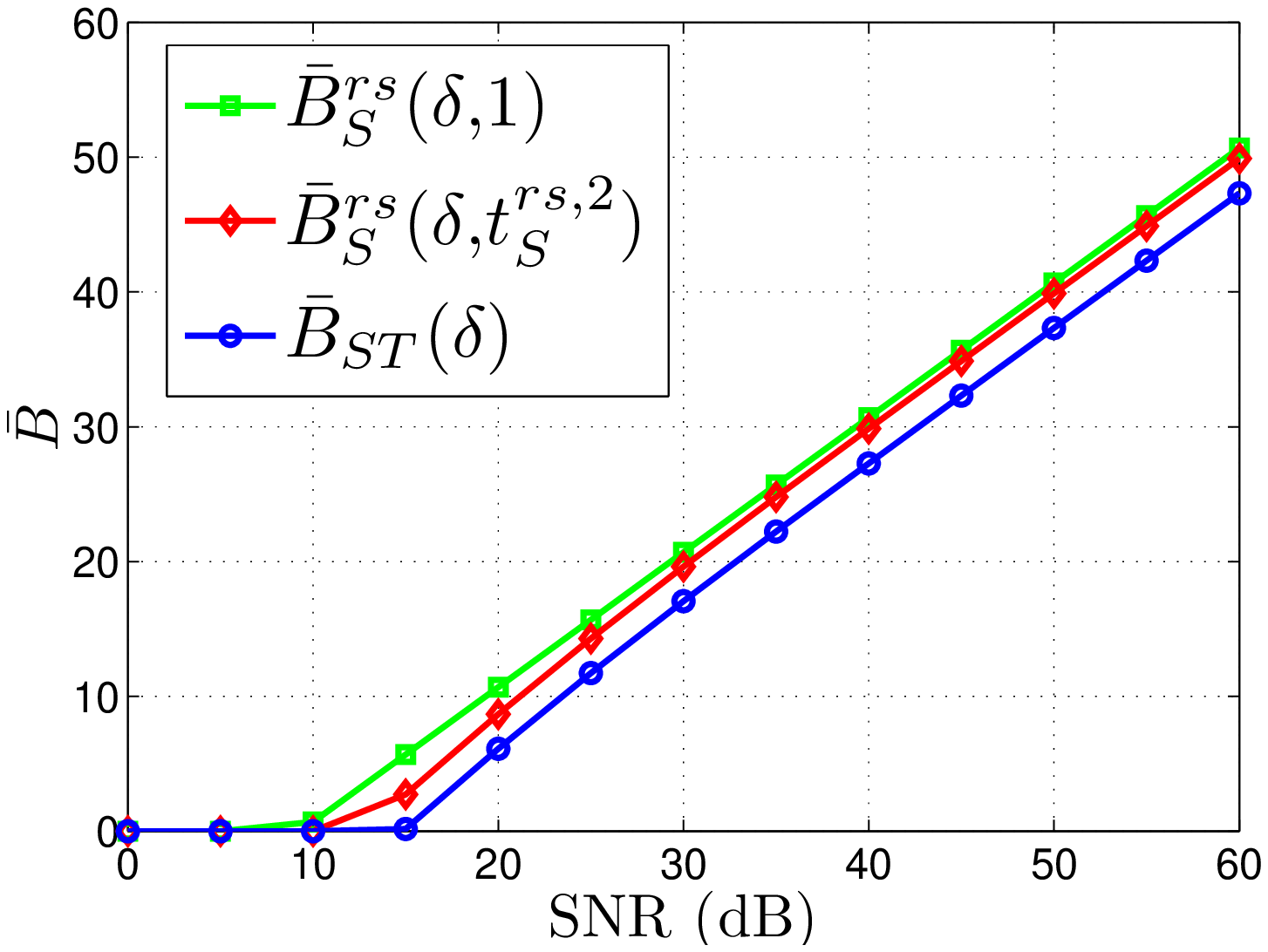}
                %\caption{$\mathcal{D}_1$}
                \label{fig:BST}
        }%\end{subfigure}
\subfigure[Sum rate]{%{0.32\textwidth}
                \centering
                \includegraphics[width=0.26\textwidth,height=3.5cm]{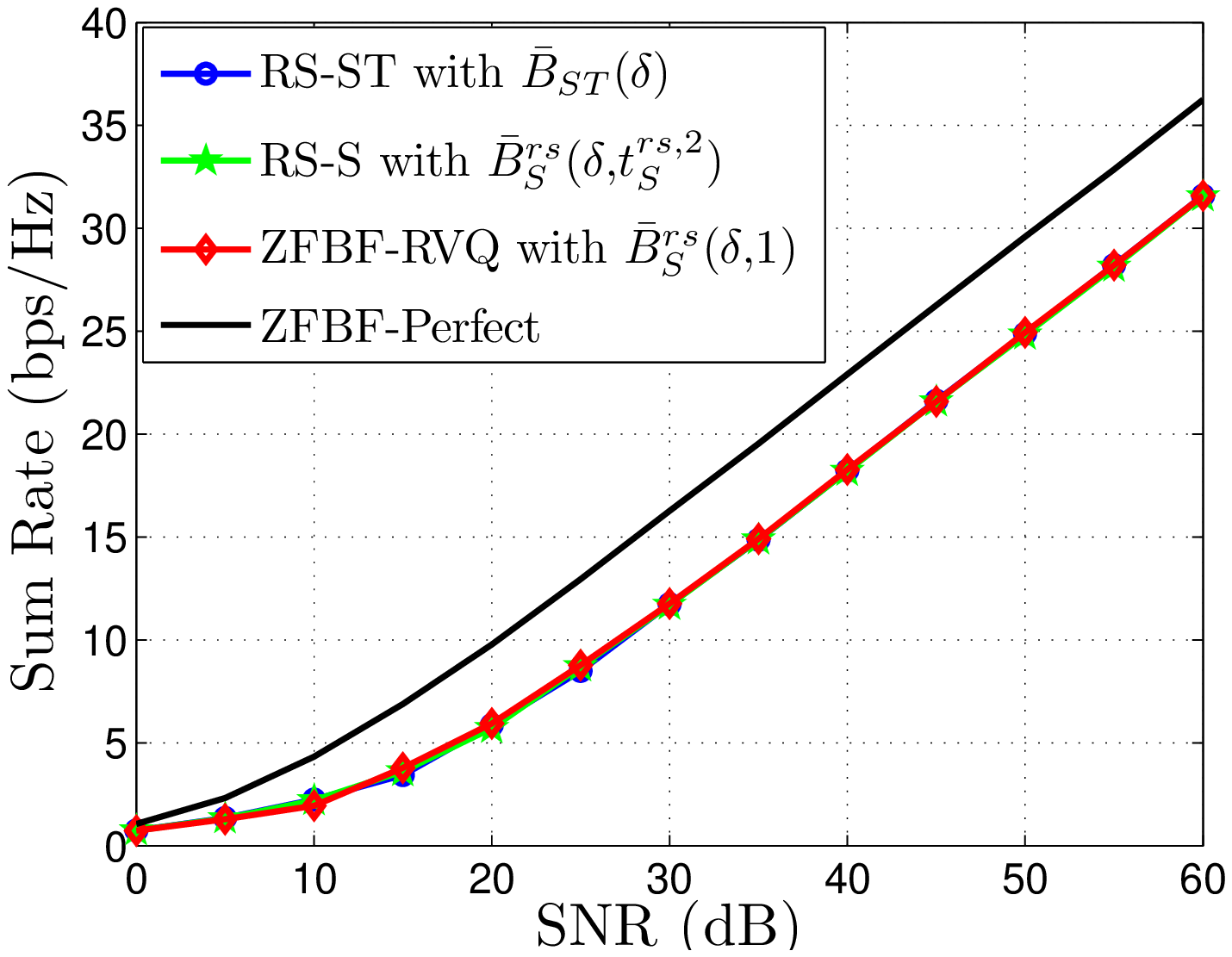}
                %\caption{$\mathcal{D}_2$}
                \label{fig:STvsS2}
        }%\end{subfigure}
\caption{$M{=}4$ and $\tau{=}14$, to achieve maximum ${\log}_2\delta{=}6$ bps/Hz rate loss}\label{fig:BdeltaST}
\end{figure}

Setting the maximum allowable rate loss to be ${\log}_2\delta{=}6$ bps/Hz, we plot $\bar{B}_S^{rs}(\delta{,}1)$, $\bar{B}_S^{rs}(\delta{,}t_S^{rs{,}2})$ and $\bar{B}_{ST}(\delta)$ in Figure \ref{fig:BST} for $M{=}4$ and $\tau{=}14$. We can see that RS-S (i.e., $\bar{B}_S^{rs}(\delta{,}t_S^{rs{,}2})$) offers roughly $1$ bit reduction compared to ZFBF with RVQ (i.e., $\bar{B}_S^{rs}(\delta{,}1)$) at high SNR, which is smaller than the overhead reduction shown in Figure \ref{fig:BM4b64} with $\tau{=}0$. In addition, RS-ST (i.e., $\bar{B}_{ST}(\delta)$) enables $1{\sim}2$ bits reduction over RS-S (i.e., $\bar{B}_S^{rs}(\delta{,}t_S^{rs{,}2})$).

Figure \ref{fig:STvsS2} illustrates the simulation result of the sum rate performance achieved by applying $\bar{B}_S^{rs}(\delta{,}1)$ in \eqref{eq:B2} to ZFBF with RVQ, $\bar{B}_S^{rs}(\delta{,}t_S^{rs{,}2})$ in \eqref{eq:B2} to RS-S and $\bar{B}_{ST}(\delta)$ in \eqref{eq:BST} to RS-ST. Besides, for RS-S, the simulation is carried out with the power splitting ratio $t_S^{rs{,}2}$ in \eqref{eq:t4}, and for RS-ST, the simulation is carried out with the power splitting ratios $t_{ST{,}\alpha}$ and $t_{ST{,}\beta}$ in \eqref{eq:tab}. We can see that 1) all the aforementioned schemes achieve less than $6$bps/Hz rate loss relative to ZFBF with perfect CSIT with their respective scaling laws of $\bar{B}$, and 2) all the schemes achieve almost the same sum rate performance. This confirms the feedback overhead reduction benefits stated in Remark \ref{rmk:BSJMB2} and Eq.\eqref{eq:BSTminusBS}. 

\section{Performance Comparison}\label{sec:performance}
\begin{figure}[t]
\renewcommand{\captionfont}{\small}
\captionstyle{center}
\centering \subfigure[RS-S vs. SU/MU, $M{=}4$.]{%{0.32\textwidth}
                \centering
                \includegraphics[width=0.3\textwidth,height=4cm]{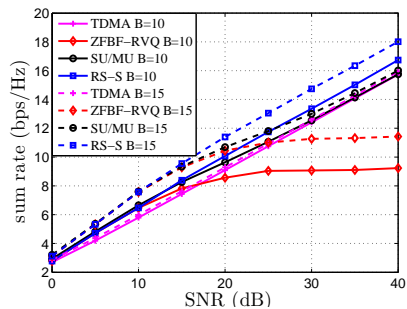}
                %\caption{$\mathcal{D}_1$}
                \label{fig:compare1}
        }\\%\end{subfigure}
        \subfigure[RS-S, RS-ST vs. SU/MU, $M{=}4$, $\tau{=}18$]{%{0.32\textwidth}
                \centering
                \includegraphics[width=0.3\textwidth,height=4cm]{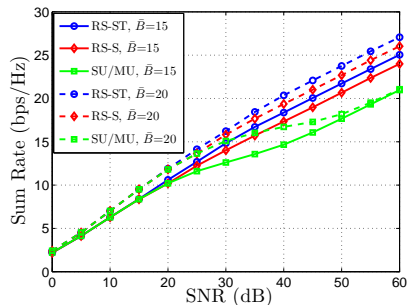}
                %\caption{$\mathcal{D}_1$}
                \label{fig:compare2}
        }
\caption{Sum rate performances of RS-ST, RS-S and SU/MU mode switching.}\label{fig:DP2}
\end{figure}
According to the power splitting ratios discussed in the previous analysis, when the number of feedback bits is fixed, both RS-S and RS-ST schemes perform similarly to ZFBF with RVQ at low and medium SNR, while they transmit common messages with most of the power at high SNR such that the sum rate is increasing rather than saturating. A counterpart of this kind of transmission is SU/MU switching, which dynamically switches between ZFBF with RVQ (i.e., multiuser mode) and TDMA (i.e., single-user mode) to maximize the sum rate. At high SNR, since the sum rate achieved with ZFBF with RVQ saturates, TDMA dominate SU/MU and benefits from the multiuser diversity. Thus, we can wonder whether RS-S and RS-ST schemes still outperform SU/MU switching in terms of sum rate when the number of feedback bits is fixed. This leads to the discussion of this section.

Note that RS is a general approach that integrates common messages on top of conventional multiuser transmission to enhance the sum rate performance with imperfect CSIT. Previous sections consider a simple case and aim to characterize the fundamental benefit of transmitting common messages in terms of sum rate performance. In this section, to gain more insights into the RS approach, we focus on a different design of RS-S and RS-ST schemes compared to the discussions in the previous sections. In particular, given the framework of RS-S in \eqref{eq:generalP1s} and RS-ST in \eqref{eq:s1STJMB} and \eqref{eq:s2STJMB}, we choose
\begin{itemize}
\item $\mathbf{w}_{cl}$ is chosen as the dominant right-singular vector of $\hat{\mathbf{H}}_l{\triangleq}[\hat{\mathbf{h}}_{1l}{,}\hat{\mathbf{h}}_{2l}]^H{,}l{=}1{,}2$;
\item $\mathbf{w}_{kl}{=}\frac{\mathbf{p}_{kl}}{{\parallel}\mathbf{p}_{kl}{\parallel}}$ is the ZF-pecoder, $k{=}1{,}2$, where $\left[\mathbf{p}_{1l}{,}\mathbf{p}_{2l}\right]{=}\hat{\mathbf{H}}_l^\dagger$, $l{=}1{,}2$.
\item In the RS-ST scheme, we choose $\mathbf{w}_{01}{=}\mathbf{w}_{11}$ and $\mathbf{w}_{02}{=}\mathbf{w}_{22}$;
\item In the RS-S scheme, the power splitting ratio $t$ is given by \eqref{eq:t} when the two receivers have equal feedback qualities, while it is chosen as \eqref{eq:t3} in the scenario with alternating receiver-specific feedback qualities. In the RS-ST scheme, $t_{\alpha}$ and $t_{\beta}$ are chosen as \eqref{eq:tab}.
\end{itemize}

In the simulation, we compute the ergodic rate of SU/MU switching as $R^{SU/MU}{\triangleq}\frac{1}{2}(R_1^{SU/MU}{+}R_2^{SU/MU})$, where $R_l^{SU/MU}{\triangleq}\mathbb{E}\left[{\max}\left(R_l^{TDMA}{,}R_l^{ZFBF}\right)\right]$ refers to the rate achieved in channel use $l$, $l{=}1{,}2$. Note that the precoders employed in the ZFBF with RVQ are the same as $\mathbf{w}_{kl}$ in the RS-S and RS-ST scheme and $R_l^{TDMA}{\triangleq}{\log}_2(1{+}P\max_{k{=}1{,}2}|\mathbf{h}_{kl}^H\hat{\mathbf{h}}_{kl}|^2)$.

Figure \ref{fig:compare1} compares the sum rate achieved with RS-S scheme and SU/MU in the scenario where the two receivers have equal feedback qualities. As shown, although RS-S offers no DoF gain compared to SU/MU, it still enables a SNR gain over SU/MU, for $B{=}10$, $15$ and $M{=}4$. Intuitively, the reason can be drawn from the power splitting ratio $t_S^{eq}$ in \eqref{eq:t}, though the precoders considered in this section is slightly different. For a fixed value of $B$, the power splitting ratio $t_S^{eq}$ in \eqref{eq:t} tends to zero at high SNR, but the amount of power that is allocated to the private messages, i.e., $Pt_S^{eq}$, remains to be a constant, namely $Pt_S^{eq}\stackrel{P{\to}\infty}{=}\frac{2(M{-}1)}{M}2^{\frac{B}{M{-}1}}$. Then, we can see that the rate of the common message is limited by the receiver with the weakest effective channel gain and is probably lower than the rate of the single message sent via SU/MU (SU/MU boils down to TDMA at high SNR for fixed B). But the contribution of the rates of the private and common messages altogether leads to a higher sum rate than SU/MU, if the feedback quality is good enough. Moreover, such a rate gap offered by RS-S over SU/MU increases with $B$.

Similar observations can be seen from Figure \ref{fig:compare2}, which illustrates the sum rate performance of the RS-S, RS-ST schemes and SU/MU in the scenario with alternating receiver-specific feedback qualities. As shown, both schemes yield a significant SNR gain over SU/MU. Besides, RS-ST offers about $3$dB SNR gain over RS-S scheme when $\tau{=}18$.

As pointed out earlier, RS is a general approach that integrates common messages on top of conventional multiuser transmissions, such as Regularized-ZF, THP, etc, in order to enhance the sum rate achieved by simply performing conventional multiuser transmissions in the presence of imperfect CSIT. Although we have not been able to perform an analysis of the RS-S and RS-ST with the precoders specified in this section, the benefit of transmitting the common message found in this paper would be extendable to RS with any conventional multiuser transmissions. 

\section{Conclusion}\label{sec:conclusion}
In this paper, focusing on a two-receiver MISO BC with quantized CSIT (based on RVQ), we investigate the ergodic sum rate of two new multiuser transmission schemes based on a rate-splitting strategy, known as RS-S and RS-ST. In these two schemes, the message of one receiver is divided into a common and a private part, where the private messages are transmitted via ZFBF using a fraction of the total power, while the common messages are transmitted via a space design in the RS-S scheme and via a space-time design in the RS-ST scheme using the remaining power. We derive an upper-bound on the sum rate loss incurred by each scheme relative to ZFBF with perfect CSIT, which highlights that an increase in the number of feedback bits leads to a SNR/rate offset of the sum-rate performance. This gain is higher than that obtained by single-user transmission, i.e., TDMA, and contrasts with that of conventional multiuser transmission, i.e., ZFBF with quantized CSIT, where the sum rate saturates at high SNR. Besides, RS-ST scheme outperforms RS-S scheme by a constant gap in the scenario with alternating receiver-specific feedback qualities. A scaling law of the feedback overhead required to achieve a certain maximum allowable rate loss is derived for each scheme. It shows that RS-S scheme enables a feedback overhead reduction compared with ZFBF with quantized CSIT, and RS-ST scheme offers a further reduction over RS-S scheme in the case of alternating receiver-specific feedback qualities. At last, through simulation, we show that both schemes offer a significant SNR gain over SU/MU switching at high SNR. Those results provide fundamental insights into the benefit of the RS approach in the presence of imperfect CSIT, and would be extendable to RS approach with other type of private message transmission. Besides, it is expected that such a rate splitting strategy will have fundamental impacts on various MIMO wireless network configurations where the performance is limited by inaccurate CSIT (e.g., K-user Broadcast and Interference channel, massive MIMO) and lead to novel transmission strategies for beyond LTE-A that do not only rely on conventional SU/MU switching.

\section*{Appendix}
\subsection{Proof of Lemma \ref{lemma:jcdf}}
Since $(X_{11}{,}X_{12})$ are statistical equivalent with $(X_{21}{,}X_{22})$, let us drop the index $k$ and introduce $\beta_1{=}|\bar{\mathbf{h}}\mathbf{w}_c|^2$, $\beta_2{=}|\bar{\mathbf{h}}\mathbf{w}|^2$ and $a{=}{\parallel}\mathbf{h}{\parallel}^2$. Then, we have $X_1{=}\beta_1a$ and $X_2{=}\beta_2a$. As $\mathbf{h}$ is independent of $\mathbf{w}_c$ and $\mathbf{w}$, $\beta_1$ and $\beta_2$ are beta $(1{,}M{-}1)$ random variables. Besides, $a\stackrel{d}{\sim}\chi^2(M)$ is independent of $\beta_1$ and $\beta_2$. The CDF of $\beta_i{,}i{=}1{,}2$ and the PDF of $a$ are given by{\small
\begin{equation}
F_{\beta_i}(\beta_i){=}\left\{\begin{array}{ll}1{-}(1{-}\beta_i)^{M{-}1}&\beta_i{\leq}1\\1&\beta_i{>}1\end{array}\right.,\quad
f_A(a){=}\frac{1}{\Gamma(M)}a^{M{-}1}e^{{-}a},
\end{equation}}respectively. Denoting $x_1^\prime{\triangleq}\min\{x_1{,}x_2\}$ and $x_2^\prime{\triangleq}\max\{x_1{,}x_2\}$, the joint CDF of $X_1$ and $X_2$ are derived as $F(x_1{,}x_2){=}\int_0^{\infty}Pr(X_1{\leq}x_1|A{=}a)Pr(X_2{\leq}x_2|A{=}a)f_A(a)da$ due to the fact that $X_1$ and $X_2$ are independent conditioned on $A{=}a$, i.e., $\beta_1$ and $\beta_2$ are independent according to the analysis in \cite{Jin06}. Then, replacing $Pr(X_i{\leq}x_i|A{=}a)$ by $F_{\beta_i}(\frac{x_i}{a})$, $Pr(X_1{\leq}x_1{,}X_2{\leq}x_2)$ is further derived as
{\small\begin{align}
&\int_0^{x_1^\prime}f(a)da{+}
\int_{x_1^\prime}^{x_2^\prime}(1{-}(1{-}\frac{x_1^\prime}{a})^{M{-}1})f_A(a)da\nonumber\\
&{+}\int_{x_2^\prime}^{\infty}(1{-}(1{-}\frac{x_1}{a})^{M{-}1})(1{-}(1{-}\frac{x_2}{a})^{M{-}1})f_A(a)da\nonumber\\
{=}&\int_0^\infty f_A(a)da{-}\sum_{i{=}1}^2\int_{x_i^\prime}^\infty(1{-}\frac{x_i^\prime}{a})^{M{-}1}f_A(a)da\nonumber\\
&{+}\int_{x_2^\prime}^{\infty}(1{-}\frac{x_1^\prime}{a})^{M{-}1}(1{-}\frac{x_2^\prime}{a})^{M{-}1}f_A(a)da,
\label{eq:Pr0}
\end{align}}In \eqref{eq:Pr0}, it is straight forward that the first term is $1$. Simply replacing $a{=}\tilde{a}{+}x_i^\prime$ in the second term, one has $\int_0^{\infty}\frac{\tilde{a}e^{-x_i^{\prime}{-}\tilde{a}}}{\Gamma(M)}d\tilde{a}{=}e^{-x_i^\prime}$ for $i{=}1{,}2$. Let $\xi(x_1{,}x_2)$ denote the last term. It can be derived as
{\small\begin{align}
\xi(x_1{,}x_2){=}&\frac{1}{\Gamma(M)}\int_{x_2^\prime}^\infty\sum_{i{=}0}^{M{-}1}\sum_{j{=}0}^{M{-}1}{{M{-}1}\choose{i}}{{M{-}1}\choose{j}}
{\times}\nonumber\\&(-x_1)^{M{-}1{-}i}(-x_2)^{M{-}1{-}j}a^{i{+}j{+}1{-}M}e^{-a}da\nonumber\\
{=}&\frac{1}{\Gamma(M)}\sum_{i{=}0}^{M{-}1}\sum_{j{=}0}^{M{-}1}{{M{-}1}\choose{i}}{{M{-}1}\choose{j}}{\times}\nonumber\\
&(-x_1)^{M{-}1{-}i}(-x_2)^{M{-}1{-}j}\Gamma(i{+}j{+}2{-}M{,}x_2^\prime).
\end{align}}Consequently, \eqref{eq:xi} and \eqref{eq:jcdf} are immediate. $\hfill\Box$

\subsection{Proof of Lemma \ref{lemma:ElogLB}}\label{app:lemmaElog}
The expectation of $Z$ with support $({-}\infty{,}\infty)$ writes as $\mathbb{E}[Z]{=}\int_{-\infty}^\infty z dF_Z(z)$. It can be derived as
{\small\begin{align}
&\int_{-\infty}^0z dF_Z(z){+}\int_0^\infty zdF_Z(z)\nonumber\\
{=}&-\int_{-\infty}^0\int_{z}^01d\theta dF_Z(z){+}\int_0^\infty\int_0^z1d\theta dF_Z(z)\\
{=}&-\int_{-\infty}^0\int_{-\infty}^\theta dF_Z(z)d\theta{+}\int_0^\infty\int_z^\infty dF_Z(z)d\theta\\
{=}&-\int_{-\infty}^0(F_Z(\theta){-}F_Z(-\infty))d\theta{+}\int_0^\infty(F_Z(\infty){-}F_Z(\theta))d\theta\\
{=}&-\int_{-\infty}^0F_Z(\theta)d\theta{+}\int_0^\infty(1{-}F_Z(\theta))d\theta.\label{eq:F}
\end{align}}Similarly, for $\tilde{Z}$ with the support $({-}\infty{,}\infty)$, we can write $\mathbb{E}[\tilde{Z}]{=}\int_0^\infty(1{-}F_{\tilde{Z}}(\theta))d\theta{-}\int_{-\infty}^0F_{\tilde{Z}}(\theta)d\theta$. Since $F_Z(z){\leq}F_{\tilde{Z}}(z)$, we have $\mathbb{E}[Z]{\geq}\mathbb{E}[\tilde{Z}]$. $\hfill\Box$

\subsection{Proof of Proposition \ref{theo:RateLossP1}}
Apparently, $R_1^p$ and $R_1(t)$ are statistically equivalent with $R_2^p$ and $R_2(t)$ respectively, thus we only need to upper-bound $R_k^p{-}R_k(t)$ and lower-bound $R_c(t)$. Specifically,
{\small\begin{IEEEeqnarray}{rcl}
\!\!\!\!&R_k^p&{-}R_k(t)\nonumber\\
\!\!\!\!&{\leq}&\mathbb{E}\left[{\log_2}(1{+}|\mathbf{h}_k^H\mathbf{w}_{k{,}pf}|^2\frac{P}{2})\right]{-}
\mathbb{E}\left[{\log_2}(1{+}|\mathbf{h}_k^H\mathbf{w}_k|^2\frac{Pt}{2})\right]{+}\nonumber\\
\!\!\!\!&&\mathbb{E}\left[{\log_2}(1{+}|\mathbf{h}_k^H\mathbf{w}_j|^2\frac{Pt}{2})\right]\IEEEyessubnumber\label{eq:phifunc0}\\
\!\!\!\!&{=}&\frac{1}{\ln 2}\left[\phi(\frac{P}{2}){-}\phi(\frac{Pt}{2})\right]{+}
\mathbb{E}\left[{\log_2}(1{+}|\mathbf{h}_k^H\mathbf{w}_j|^2\frac{Pt}{2})\right]\IEEEyessubnumber\label{eq:phifunc}\\
\!\!\!\!&{\leq}&\frac{1}{\ln 2}\left[\phi(\frac{P}{2}){-}\phi(\frac{Pt}{2})\right]\!{+}
{\log_2}\!\left(\!1{+}\frac{Pt}{2}\mathbb{E}\left[{\parallel}\mathbf{h}_k{\parallel}^2|\bar{\mathbf{h}}_k^H\mathbf{w}_j|^2\right]\!\right)
\IEEEyessubnumber\label{eq:jensen}\\
\!\!\!\!&{\leq}&\frac{1}{\ln 2}\left[\phi(\frac{P}{2}){-}\phi(\frac{Pt}{2})\right]{+}
{\log_2}\left(1{+}\frac{Pt M}{2(M{-}1)}2^{-\frac{B}{M{-}1}}\right).\IEEEyessubnumber\label{eq:RVQerror}
\end{IEEEeqnarray}}As $\mathbf{w}_{k{,}pf}$ is a unit-norm vector randomly chosen from the $M{-}1$-dimensional null space of $\mathbf{h}_j{,}k{\neq}j$, the random variable $|\mathbf{h}_k^H\mathbf{w}_{k{,}pf}|^2$ is exponential distributed with parameter $1$. Thus, the first two terms in \eqref{eq:phifunc0} have the same form $\mathbb{E}_r[{\log_2}(1{+}ar)]$. \eqref{eq:phifunc} is obtained by calculating the integral $\int_0^\infty{\log_2}(1{+}ar)e^{-r}dr{=}\frac{\phi(a)}{{\ln}2}$. \eqref{eq:jensen} follows Jensen's Inequality. \eqref{eq:RVQerror} is obtained by Lemma \ref{lemma:upperQuantizeError}, and the fact that ${\parallel}\mathbf{h}_k{\parallel^2}\stackrel{d}{\sim}\chi^2(M)$.

Recalling that $Y{=}{\min}(Y_1{,}Y_2)$ where $Y_k{=}\frac{|\mathbf{h}_k^H\mathbf{w}_c|^2}{1{+}|\mathbf{h}_k^H\mathbf{w}_k|^2\frac{Pt}{2}}$ and using Jensen's Inequality, we lower-bound $R_c$ as $R_c(t){\geq}{\log}_2\left(1{+}P(1{-}t)e^{\mathbb{E}\left[{\ln}Y\right]}\right)$ due to the fact that ${\log_2}(1{+}ae^r)$ is a convex function of $r$. Since the r.h.s. of \eqref{eq:cdfy} gives an approximate upper-bound of $F_Y(y)$ according to Lemma \ref{lemma:CDFUB}, we can use r.h.s. of \eqref{eq:cdfy} to lower-bound $\mathbb{E}\left[{\ln}Y\right]$ following Lemma \ref{lemma:ElogLB}. Consequently, one has
{\small\begin{align}
\mathbb{E}[{\ln}Y]{\geq}&\int_0^\infty \left(\frac{Pt}{(1{+}\frac{Pt}{2}y)^3}e^{-2y}{+}\frac{2}{(1{+}\frac{Pt}{2}y)^2}e^{-2y}\right)\cdot{\ln}y\,dy\nonumber\\
{=}&\left(\frac{4}{Pt}{-}1\right)\phi(\frac{Pt}{4}){-}\gamma{-}{\ln}2{-}1,\label{eq:Elny}
\end{align}}where $\frac{Pt}{(1{+}\frac{Pt}{2}y)^3}e^{-2y}{+}\frac{2}{(1{+}\frac{Pt}{2}y)^2}e^{-2y}$ is obtained by calculating the derivative of the r.h.s. of \eqref{eq:cdfy}. Combining \eqref{eq:RVQerror} and \eqref{eq:Elny} yield \eqref{eq:dRt}.

Note that the proof of Proposition \ref{prop:ratelossQ1} follows similarly. The only difference lies in \eqref{eq:RVQerror}, where the last term is upper-bounded by the receiver-specific feedback quality, i.e., $B_{kl}$. $\hfill\Box$

\subsection{Proof of Corollary \ref{coro:BSrs} and Derivation of \eqref{eq:t4}}
In \eqref{eq:RLQ1}, one can write ${\log}_2(1{+}t\Lambda_\alpha){+}{\log}_2(1{+}t\Lambda_\beta){=}{\log}_2\left[\left(t\eta{-}1\right)^2{+}t\eta\Theta\right]$,
where $\eta{=}\sqrt{\Lambda_\alpha\Lambda_\beta}$ and $\Theta{=}2{+}\frac{\sqrt{\Lambda_{\alpha}}}{\sqrt{\Lambda_{\beta}}}{+}\frac{\sqrt{\Lambda_{\alpha}}}{\sqrt{\Lambda_{\beta}}}{=} 2^{\frac{-\tau}{2(M{-}1)}}{+}2^{\frac{\tau}{2(M{-}1)}}{+}2$ are functions of $\bar{B}$ and $\tau$, respectively. Setting $\Delta\tilde{R}_{S}^{rs}{=}{\log}_2\delta$ and inverting it w.r.t. $\eta$, one has $(\eta{-}\frac{1}{t})^2{+}\frac{\eta\Theta}{t}{=}\frac{\delta\left(
1{+}\frac{P(1{-}t)}{2}e^{\kappa(t)}\right)}
{t^2\cdot2^{2\epsilon(t)}}$. Solving this quadratic formula yields
{\small\begin{align}
\eta{=}&\sqrt{\frac{\Theta^2{-}4\Theta}{4t^2}{+}\frac{\delta\left(
1{+}\frac{P(1{-}t)}{2e^{1{+}\gamma}}e^{(\frac{4}{Pt}{-}1)\phi(\frac{Pt}{4})}\right)}
{t^2\cdot2^{\frac{2}{\ln2}\left[\phi(\frac{P}{2}){-}\phi(\frac{Pt}{2})\right]}}}{-}\frac{\Theta{-}2}{2t}.
\end{align}}Proposition \ref{coro:BSrs} is immediate by expanding $\eta$ as a function of $\bar{B}$.$\hfill\Box$

Next, to calculate $t_S^{rs{,}2}{=}\arg\min_{0{<}t{\leq}1}\bar{B}_S^{rs}(\delta{,}t)|_{P{\to}\infty}$, let us replace $\frac{1}{t}$ with $r$ and write $\bar{B}_S^{rs}(\delta{,}r)|_{P{\to}\infty}$ and $\frac{d\bar{B}_S^{rs}(\delta{,}r)|_{P{\to}\infty}}{dr}$ as
{\small\begin{IEEEeqnarray}{rcl}
\bar{B}_S^{rs}(\delta{,}r)|_{P{\to}\infty}&{=}&\sqrt{\frac{\Theta^2{-}4\Theta}{4}r^2{+}\delta(1{+}\frac{2}{e}r{-}\frac{2}{e})}{-}
\frac{\Theta{-}2}{2}r,\IEEEyessubnumber\\
\frac{d\bar{B}_S^{rs}(\delta{,}r)|_{P{\to}\infty}}{dr}&{=}&\frac{\frac{\Theta^2{-}4\Theta}{2}r{+}\frac{2}{e}\delta}
{2\sqrt{\frac{\Theta^2{-}4\Theta}{4}r^2{+}\delta(1{+}\frac{2}{e}r{-}\frac{2}{e})}}{-}\frac{\Theta{-}2}{2}.\IEEEyessubnumber\label{eq:derivative1}
\end{IEEEeqnarray}}By setting $\frac{d\bar{B}_S^{rs}(\delta{,}r)|_{P{\to}\infty}}{dr}{=}0$, we find that the stationary points satisfy
{\small\begin{equation}
(\Theta^2{-}4\Theta){r^*}^2{+}\frac{8\delta}{e}r^*{=}\frac{4\delta^2}{e^2}{-}(\Theta{-}2)^2\delta(1{-}\frac{2}{e}).\label{eq:stationary}
\end{equation}}When $\Theta{=}4$, i.e., $\tau{=}0$, we have $r^*{=}\frac{\delta}{2e}{+}1{-}\frac{e}{2}$, which leads to $t_S^{eq{,}2}$ in \eqref{eq:t2}. Since we consider $\tau{>}0$ (or $\Theta{>}4$), by solving the quadratic formula \eqref{eq:stationary}, we find that one stationary point writes as
{\small\begin{equation}
r^*{=}\sqrt{\frac{4(\Theta{-}2)^2}{e^2(\Theta^2{-}4\Theta)^2}\delta^2
{-}\frac{(\Theta{-}2)^2}{\Theta^2{-}4\Theta}\delta(1{-}\frac{2}{e})}{-}\frac{4\delta}{e(\Theta^2{-}4\Theta)}. \label{eq:rstar}
\end{equation}}It can be seen that $r^*$ in \eqref{eq:rstar} minimizes $\bar{B}_S^{rs}(\delta{,}r)|_{P{\to}\infty}$ because $\frac{d\bar{B}_S^{rs}(\delta{,}\frac{1}{r})|_{P{\to}\infty}}{dr}{>}0{,}{\forall}r{>}r^*$ and $\frac{d\bar{B}_S^{rs}(\delta{,}\frac{1}{r})|_{P{\to}\infty}}{dr}{<}0{,}\\{\forall}0{<}r{<}r^*$. Consequently, we choose $t_S^{rs{,}2}{=}\min\{1{,}\frac{1}{r^*}\}$. Then, evaluating $r^*$ in \eqref{eq:rstar} leads to the closed-form of $t_S^{rs{,}2}$ in \eqref{eq:t4} and the threshold $\delta_0(\Theta)$. $\hfill\Box$

\subsection{Proof of Proposition \ref{prop:RateLossP2}}
In the RS-ST scheme, we aim to upper-bound $R_1^p{-}R_{u11}(t_\beta{,}t_\alpha)$ and $R_2^p{-}R_{u21}(t_\beta{,}t_\alpha)$ while find lower bounds for $R_{c1}(t_\beta{,}t_\alpha)$, $R_{c2}(t_\beta{,}t_\alpha)$ and $R_{c0}(t_\beta{,}t_\alpha)$. Specifically,
{\small \begin{align}
R_1^p&{-}R_{u11}(t_\beta{,}t_\alpha)\nonumber\\
{=}&\mathbb{E}\left[{\log_2}(1{+}|\mathbf{h}_{11}^H\mathbf{w}_{11{,}opt}|^2\frac{P}{2})\right]{-}
\mathbb{E}\left[{\log_2}(1{+}\frac{|\mathbf{h}_{11}^H\mathbf{w}_{11}|^2\frac{Pt_\alpha}{2}}
{1{+}|\mathbf{h}_{11}^H\mathbf{w}_{21}|^2\frac{Pt_\beta}{2}})\right]\\
{=}&\frac{1}{\ln2}\left[\phi(\frac{P}{2}){-}\phi(\frac{Pt_\alpha}{2})\right]{+}
{\log_2}\left(1{+}\frac{Pt_\beta M}{2(M{-}1)}2^{-\frac{B_\beta}{M{-}1}}\right),\label{eq:Rp1P2}
\end{align}}where we have used the same derivation of Proposition \ref{theo:RateLossP1}. Note that since $\mathbf{w}_{21}{\in}\hat{\mathbf{h}}_{11}^\bot$, the upper-bound is a function of $B_{11}{=}B_{\beta}$. Similarly, one has
{\small \begin{multline}
R_2^p{-}R_{u21}(t_\beta{,}t_\alpha){\leq}\frac{1}{\ln2}\left[\phi(\frac{P}{2}){-}\phi(\frac{Pt_\beta}{2})\right]{+}\\
{\log_2}\left(1{+}\frac{Pt_\alpha M}{2(M{-}1)}2^{-\frac{B_\alpha}{M{-}1}}\right).\label{eq:Rp2P2}
\end{multline}}Using Assumption \ref{asp:sinr_approx}, we have ${\rm SINR}_{c1}^{(1)}{\approx}\frac{|\mathbf{h}_{11}^H\mathbf{w}_{c1}|^2P(1{-}t_\beta)}{1{+}|\mathbf{h}_{11}^H\mathbf{w}_{11}|^2\frac{Pt_\beta}{2}}$, ${\rm SINR}_{c2}^{(1)}{\approx}\frac{|\mathbf{h}_{12}^H\mathbf{w}_{c2}|^2P(1{-}t_\beta)}{1{+}|\mathbf{h}_{12}^H\mathbf{w}_{12}|^2\frac{Pt_\beta}{2}}$ and ${\rm SINR}_{c0}^{(1)}{\approx}\frac{|\mathbf{h}_{11}^H\mathbf{w}_{11}|^2\frac{P(t_\beta{-}t_\alpha)}{2}} {1{+}|\mathbf{h}_{11}^H\mathbf{w}_{11}|^2\frac{Pt_\alpha}{2}}$, where we have also used the fact that $\mathbf{w}_{01}{=}\mathbf{w}_{11}$. Then, following the footsteps in the proof of Proposition \ref{theo:RateLossP1}, $R_{c1}(t_\beta{,}t_\alpha)$ is lower-bounded by
{\small\begin{align}
R_{c1}(t_\beta{,}t_\alpha){=}&\mathbb{E}\left[{\log_2}\left(1{+}\min_{k{=}1{,}2}\frac{|\mathbf{h}_{k1}^H\mathbf{w}_{c1}|^2P(1{-}t_\beta)}
{1{+}|\mathbf{h}_{k1}^H\mathbf{w}_{k1}|^2\frac{Pt_\beta}{2}}\right)\right]\\
{\geq}&\log_2\left[1{+}\frac{P(1{-}t_\beta)}{2e^{1{+}\gamma}}e^{(\frac{4}{Pt_\beta}{-}1)\phi(\frac{Pt_\beta}{4})}\right].\label{eq:Rc1P2}
\end{align}}The derivation of $R_{c2}(t_\beta{,}t_\alpha)$ follows similarly as it is statistically equivalent with $R_{c1}(t_\beta{,}t_\alpha)$. Then, it remains to bound $R_{c0}(t_\beta{,}t_\alpha)$, which writes as
{\small\begin{align}
R_{c0}(t_\beta{,}t_\alpha){=}&\mathbb{E}\left[\log_2\left(1{+}\min_{k{=}1{,}2{,}l{=}k}
\frac{|\mathbf{h}_{kl}^H\mathbf{w}_{kl}|^2\frac{P(t_\beta{-}t_\alpha)}{2}}
{1{+}|\mathbf{h}_{kl}^H\mathbf{w}_{kl}|^2\frac{Pt_\alpha}{2}}\right)\right]\nonumber\\
{=}&\mathbb{E}\left[\log_2\left(\frac{1{+}x\frac{Pt_\beta}{2}}{1{+}x\frac{Pt_\alpha}{2}}\right)\right],\label{eq:c0_rate1}
\end{align}}where $x{=}\min(|\mathbf{h}_{11}^H\mathbf{w}_{11}|^2{,}|\mathbf{h}_{22}^H\mathbf{w}_{22}|^2)$ and \eqref{eq:c0_rate1} is due to the fact that the function $\log_2(\frac{1{+}bx}{1{+}ax})$ is monotonically increasing with $x$ if $b{>}a$. Clearly, $x$ is the minimum of two independent exponential random variables, its CDF writes as $F(x){=}1{-}e^{-2x}$. Then, the \eqref{eq:c0_rate1} can be derived as
{\small\begin{align}
R_{c0}(t_\beta{,}t_\alpha){=}&\mathbb{E}\left[\log_2\left(1{+}x\frac{Pt_\beta}{2}\right)\right]{-}\mathbb{E}\left[\log_2\left(1{+}x\frac{Pt_\alpha}{2}\right)\right]\\
{=}&\frac{1}{\ln2}\left[2\phi(\frac{Pt_\beta}{2}){-}\phi(\frac{Pt_\beta}{4})
{-}2\phi(\frac{Pt_\alpha}{2}){+}\phi(\frac{Pt_\alpha}{4})\right].\label{eq:Rc0}
\end{align}}
Combining \eqref{eq:Rp1P2}, \eqref{eq:Rp2P2}, \eqref{eq:Rc1P2} and \eqref{eq:Rc0}, Proposition \ref{prop:RateLossP2} holds. $\hfill\Box$ 

\bibliographystyle{IEEEtran}

\bibliography{RSjnl}

\end{document}